\documentclass[aps,prl,reprint,superscriptaddress]{revtex4-2}
\usepackage{graphicx}
\usepackage{graphicx}
\usepackage{dcolumn}
\usepackage{bm}

\begin{document}


\title{Structural and Magnetic Transitions in the Planar Antiferromagnet Ba$_4$Ir$_3$O$_{10}$}


\author{Xiang Chen}
\email{xiangchen@berbeley.edu}
\affiliation{Physics Department, University of California, Berkeley, California 94720, USA}
\affiliation{Materials Science Division, Lawrence Berkeley National Lab, Berkeley, California 94720, USA}
\author{Yu He}
\affiliation{Physics Department, University of California, Berkeley, California 94720, USA}
\affiliation{Materials Science Division, Lawrence Berkeley National Lab, Berkeley, California 94720, USA}
\author{Shan Wu}
\affiliation{Physics Department, University of California, Berkeley, California 94720, USA}
\affiliation{Materials Science Division, Lawrence Berkeley National Lab, Berkeley, California 94720, USA}
\author{Yu Song}
\affiliation{Physics Department, University of California, Berkeley, California 94720, USA}
\author{Dongsheng Yuan}
\affiliation{Materials Science Division, Lawrence Berkeley National Lab, Berkeley, California 94720, USA}
\author{Edith Bourret-Courchesne}
\affiliation{Materials Science Division, Lawrence Berkeley National Lab, Berkeley, California 94720, USA}
\author{Jacob P.C. Ruff}
\affiliation{Cornell High Energy Synchrotron Source, Cornell University, Ithaca, NY 14853, USA}
\author{Zahirul Islam}
\affiliation{Advanced Photon Source, Argonne National Laboratory, Argonne, IL 60439, USA}
\author{Alex Frano}
\affiliation{Department of Physics, University of California, San Diego, CA 92093, USA}
\author{Robert J. Birgeneau}
\email{robertjb@berkeley.edu}
\affiliation{Physics Department, University of California, Berkeley, California 94720, USA}
\affiliation{Materials Science Division, Lawrence Berkeley National Lab, Berkeley, California 94720, USA}
\affiliation{Department of Materials Science and Engineering, University of California, Berkeley, California 94720, USA}


\date{\today}

\begin{abstract}

We report the structural and magnetic ground state properties of the monoclinic compound barium iridium oxide Ba$_4$Ir$_3$O$_{10}$ using a combination of resonant x-ray scattering, magnetometry, and thermodynamic techniques. Magnetic susceptibility exhibits a pronounced antiferromagnetic transition at $T_{\text{N}}$ $\approx$ 25K, a weaker anomaly at $T_{\text{S}}$ $\approx$ 142K, and strong magnetic anisotropy at all temperatures. Resonant elastic x-ray scattering experiments reveal a second order structural phase transition at $T_{\text{S}}$ and a magnetic transition at $T_{\text{N}}$. Both structural and magnetic superlattice peaks are observed at $L$ = half integer values. The magnetization anomaly at $T_{\text{S}}$ implies the presence of magneto-elastic coupling, which conceivably facilitates the symmetry lowering. Mean field critical scattering is observed above $T_{\text{S}}$. The magnetic structure of the antiferromagnetic ground state is discussed based on the measured magnetic superlattice peak intensity. Our study not only presents essential information for understanding the intertwined structural and magnetic properties in Ba$_4$Ir$_3$O$_{10}$, but also highlights the necessary ingredients for exploring novel ground states with octahedra trimers.

\end{abstract}


\maketitle


\section{Introduction}

Iridium oxides with strong spin-orbit coupling have received substantial interest in the pursuit of new states of quantum matter. Some prominent directions include high transition temperature superconductors, Weyl semimetals, and quantum spin liquids (QSLs) \cite{Witczak-Krempa2014, Rau2016, Cao2018RPP, Bertinshaw2019ARCMP, Balents2010, Zhou2017RMP, Wan2011, Wang2011PRL, Kim2015NatPhy, Yan2015PRX}. The $5d$ iridates are promising arenas for studies because of both their effective spin $J_{\text{eff}}$ = $\frac{1}{2}$ (Ir$^{4+}$) with more accentuated quantum effects, resulting from the balance of competing energy scales \cite{Kim2008, Kim2009, Jackeli2009PRL}, and abundant structural motifs, such as the well-known cuprate analogue \cite{Bertinshaw2019ARCMP} and geometrically frustrated pyrochlore and Kagome lattices \cite{Kimchi2014}.

Recently, unconventional electronic and magnetic ground states have been reported in compounds with the basic units of Ir-trimers, $i.e.$ three face-sharing IrO$_6$ octahedra, which are much less explored so far \cite{Cao_2000SSC, Nguyen_2019PRM, Cao2020_npjQM}. For instance, both charge density wave and canted antiferromagnetic order have been observed in the quasi-one-dimensional compound BaIrO$_3$ \cite{Cao_2000SSC,LagunaMarco2010PRL, Okazaki_2018PRB, Xu_2019PRB}. More intriguingly, spin liquid states with an unconventional mechanism of magnetic frustration have been proposed in iridates with similar Ir-trimers, such as Ba$_4$NbIr$_3$O$_{12}$ and Ba$_4$Ir$_3$O$_{10}$ (Ba4310) \cite{Nguyen_2019PRM, Cao2020_npjQM}. In Ba4310, a close structural analogue to BaIrO$_3$, in a recent report no magnetic order down to 0.2K was evidenced from magnetization measurements, while strong anisotropic antiferromagnetic exchange interactions (from -766 to -169 K) and an enormously large averaged frustration parameter ($f$ $\approx$ 2000) were observed. Furthermore, the existence of itinerant fermions in insulating Ba4310 was proposed from the low temperature linear component of the specific heat and thermal conductivity. Therefore, it was postulated in Ref. \cite{Cao2020_npjQM} that Ba4310 achieves a highly unusual QSL ground state through the formation of decoupled arrays of one dimensional (1D) Tomonaga-Luttinger liquids, enabled by a greatly reduced intra-trimer exchange interaction \cite{Cao2020_npjQM, Kugel2015}. This highly unique liquid-like state, however, appears to be quite fragile with respect to external perturbations, where magnetic order can emerge by simply 2$\%$ isovalent Sr doping or by applying magnetic fields during crystal synthesis \cite{Cao2020_npjQM, cao2020quest}. Together, these reports suggest that the magnetic ground state properties in Ba4310 are extremely delicate and serve as motivation for our detailed refinement using complementary tools.

\begin{figure*}[tb]
\centering
\includegraphics*[width=16.8cm]{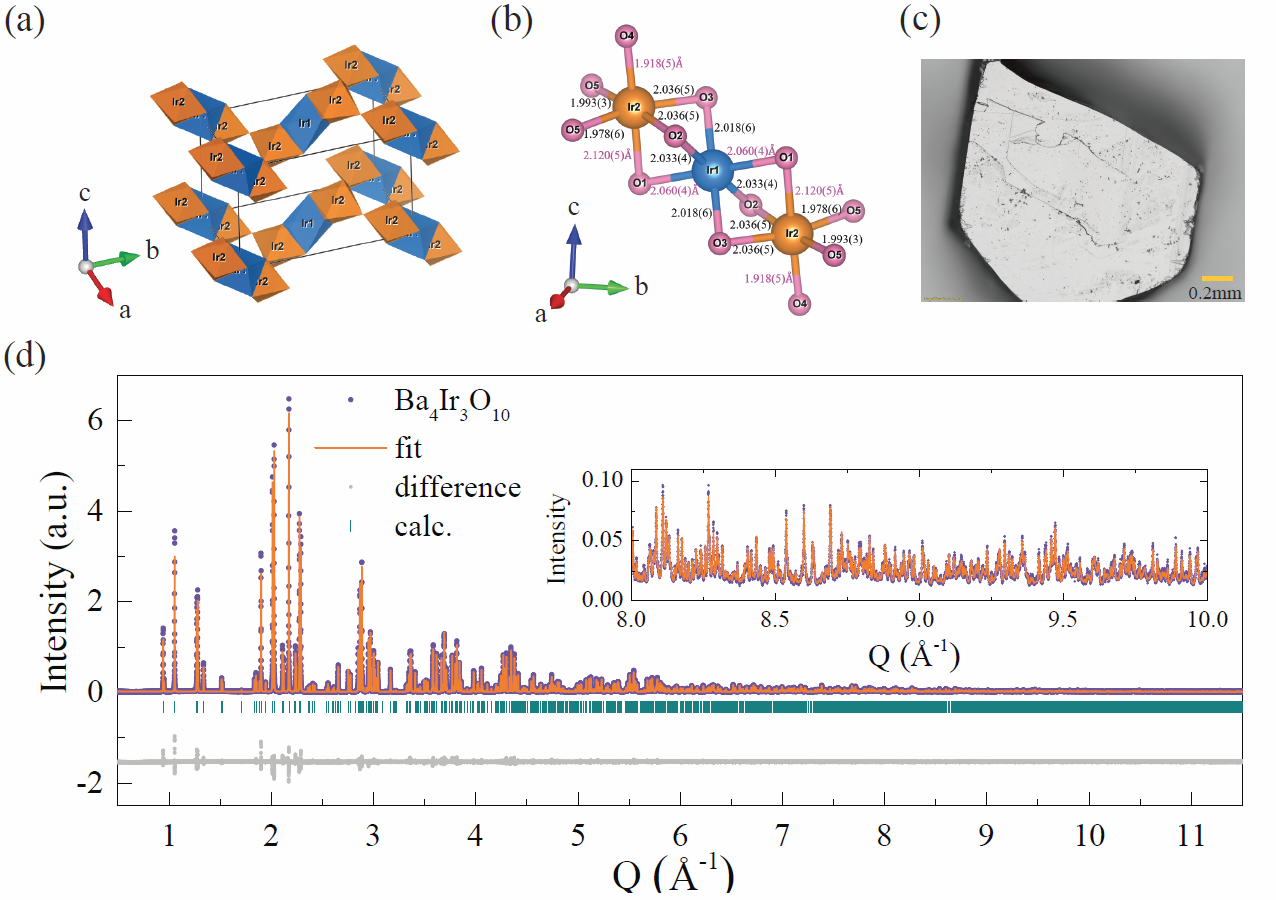}
\caption{(Color online) (a-b) Crystal structure of quasi-two dimensional Ba$_4$Ir$_3$O$_{10}$ (Ba4310) under space group P$2_1$$/a$, where the building block is the face-sharing Ir octahedra trimer as highlighted in (b). For clarity, only the Ir octahedra are shown. (b) Ir-O bond lengths, extracted from (d),  highlighted in units of \AA$\,$within an example Ir$_3$O$_{12}$ trimer. (c) Optical image of ab-plane surface of the Ba4310 single crystal for the x-ray experiments. (d) Synchrotron x-ray powder diffraction data collected at 300 K with $\lambda$ = 0.4579 \AA$\,$ on crushed Ba4310 single crystals. Data were indexed and refined to the P$2_1$$/a$ space group with $R$ factors: $\chi^2$=4.82, $R_{\text{p}}$=11.5, $R_{\text{wp}}$=12.7. Inset in panel (d) shows a magnified view of the data and fit.}
\label{fig:Fig1_alpha}
\end{figure*}

\begin{figure}[tb]
\centering
\includegraphics*[width=6.5cm]{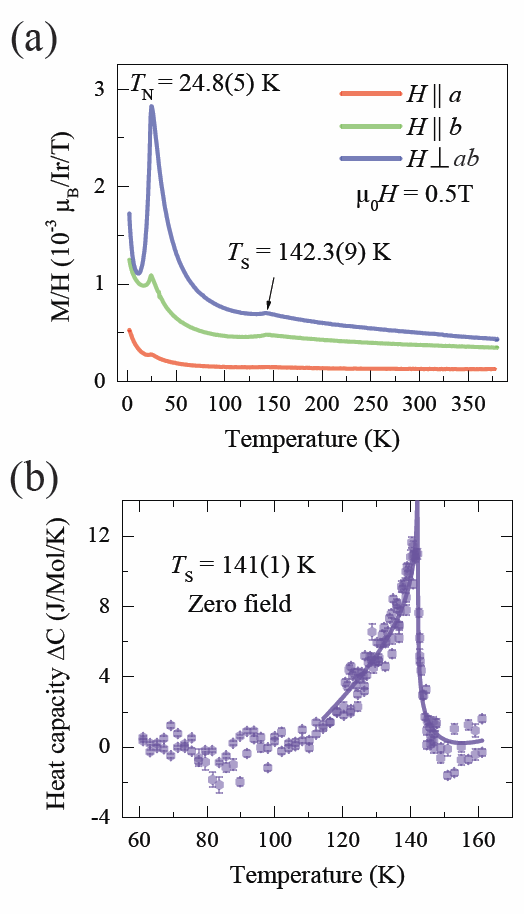}
\caption{(Color online) (a) Magnetization data of Ba4310 with external magnetic fields (0.5T) applied along different directions.
(b) Phonon background subtracted specific heat data, highlighting the phase transition at $T_{\text{S}}$ = 141(1)K. Solid line is the power law fit to data: $\Delta C=C_0+B*T+A^{\pm}{\mid}T-T_{\text{S}}{\mid}^{-\alpha}$, where $A^{\pm}$ work for $T>T_{\text{S}}$ and $T<T_{\text{S}}$, respectively. $\alpha = -0.015(1)$ is retrieved from the fit to data. Other mean field forms may also describe the data.}
\label{fig:Fig1_beta}
\end{figure}

\begin{figure}[tb]
\centering
\includegraphics*[width=6.5cm]{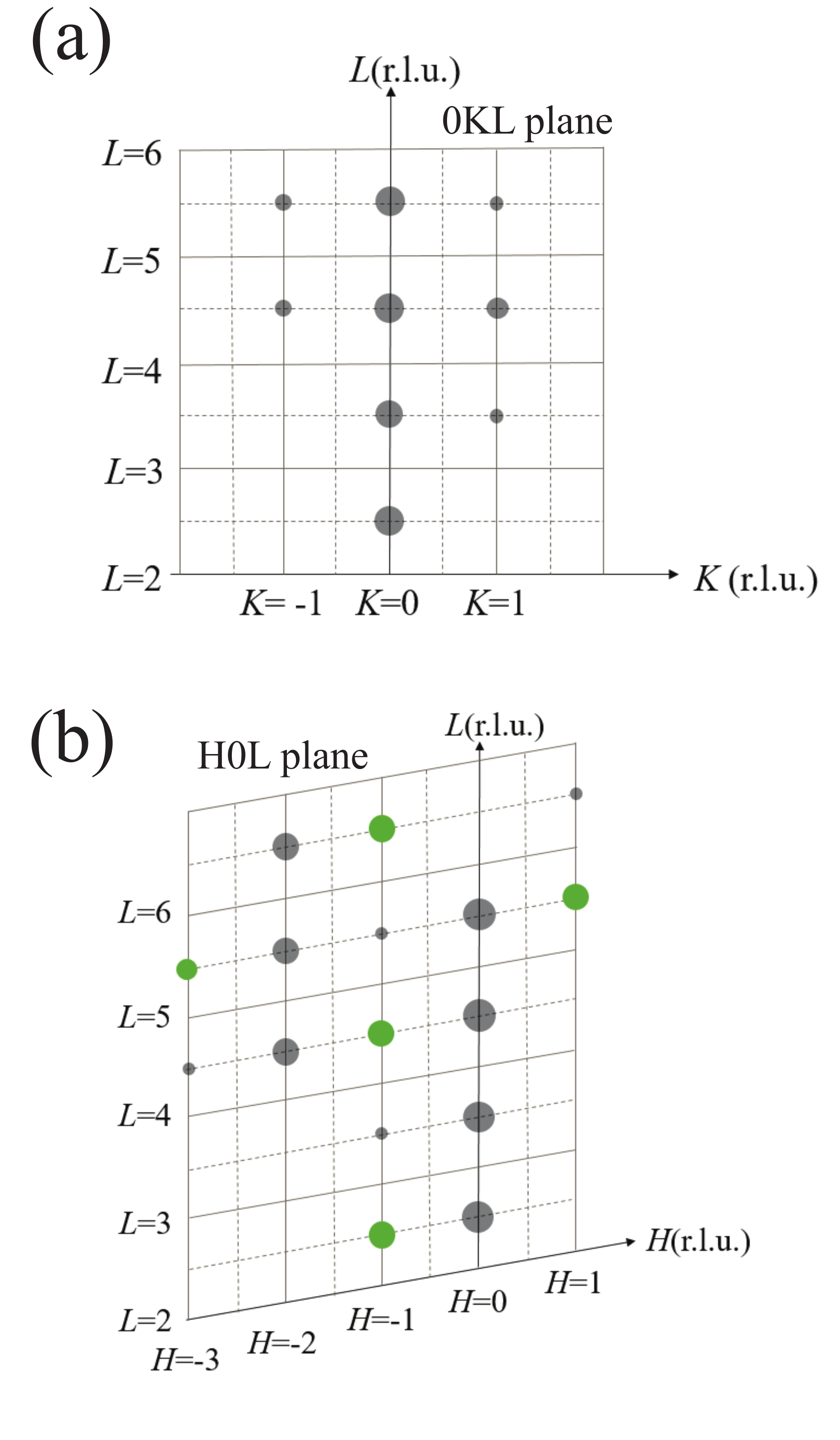}
\caption{(Color online) (a)-(b) Experimentally explored superlattice reflections are illustrated on the reciprocal 0KL ($H$=0) plane (a) and H0L ($K$=0) plane (b), respectively. Gray symbols: structural peaks; Green symbols: magnetic peaks. The size of symbols approximates the relative peak intensity.}
\label{fig:Fig1}
\end{figure}

In this work, resonant elastic X-ray scattering (REXS) \cite{Blume1985, Hannon1988PRL, Kim2009}, magnetometry and specific heat are utilized together to investigate the ground state properties of our as-grown Ba4310 single crystals. In contrast to previous results, our high quality single crystals exhibit clear magnetic order without introducing disorder by doping. Our scattering data reveal two consecutive second order structural and magnetic transitions at $T_{\text{S}}$ $\approx$ 142K and $T_{\text{N}}$ $\approx$ 25K, respectively, where both types of superlattice reflections are observed at $L$ = half integer values. Interestingly, the two transitions are consistent with the kinks in the anisotropic magnetization data, which are qualitatively similar to the data from the perturbed samples in previous report \cite{Cao2020_npjQM}.

\section{Experimental details}

Single crystals of Ba4310 were obtained using the flux growth method \cite{Chen2015PRB}. Starting materials BaCO$_3$, Ir and BaCl$_2$ powder were mixed together with a molar ratio 3: 1: 5 and placed inside a platinum (Pt) crucible. The whole mixture was heated up and held at 1400 $^\circ$C for 10 hours, then cooled down slowly at the rate of 1 $^\circ$C/min to 950 $^\circ$C, which was then furnace-cooled to room temperature. Shiny crystals were obtained by dissolving the flux with deionized water. 

The chemical composition of the crystals was verified by energy dispersive x-ray spectroscopy (EDX). The Ba : Ir ratio determined from EDX characterizations was 1.35(4), which equals within the error bars to the ideal ratio 4 : 3. The phase purity was also confirmed through powder X-ray diffraction (PXRD) on crushed single crystals, which were measured at the 11-BM beam line at the Advanced Photon Source at Argonne National Laboratory. No additional chemical phases were found within the resolution of these measurements. 

Specific heat measurements were performed with a Quantum Design Physical Property Measurement System (PPMS). Magnetization measurements were performed in a Quantum Design MPMS3 magnetometer. REXS experiments were performed on a Ba$_4$Ir$_3$O$_{10}$ single crystal (Fig. 1(c)), whose mosaicity was determined to be 0.02$^\circ$ in peak width (full width at half maximum, FWHM) on the (0 0 5) nuclear Bragg peak at $E$ = 11.215 keV. The REXS experiments were performed at the QM$2$ beamlines at the Cornell High Energy Synchrotron Source, and the 6-ID-B beamline at the Advanced Photon Source at Argonne National Laboratory. The same sample was mounted on the top of a Cu post and secured with GE varnish. A vertical scattering geometry was used with the sample aligned in the (H0L) or (0KL) scattering planes, where $\textbf{Q}$ = $(H\cdot\frac{2\pi}{a}, K\cdot\frac{2\pi}{b}, L\cdot\frac{2\pi}{c})$ is defined in reciprocal lattice units ($r.l.u.$) with lattice parameters $a$, $b$ and $c$. The data were collected near the Ir $L_3$ edge with energy $E\sim$11.215 keV, to enhance the magnetic scattering signal. Similar protocols have been extensively practiced in 5$d$ iridates \cite{Kim2009, Fujiyama2012PRL, Boseggia2013JPCM, Clancy2014PRB, HwanChun2015NatPhy, Hogan2015PRL, Chen2018NatComm}. At the QM$2$ beamline, the scattered photon energy and polarization were analyzed using silicon (4, 4, 4), and collected using a small area detector; while at 6-ID-B beamline the (0, 0, 8) reflection from a flat Highly oriented pyrolytic graphite (HOPG) analyzer crystal was utilized.

The scattering data from the REXS experiments are analyzed with one-dimensional (1D) fitting along the chosen scanned direction in reciprocal space. Below the transition temperature and away from the dynamical fluctuations, the nuclear or magnetic Bragg peak profiles are resolution-limited and empirically well described with a Lorentzian form \cite{Harris1995_PRB}:

\begin{equation}
I(\textbf{Q}) = \frac{I_0}{\pi}\frac{\kappa_0}{(\textbf{Q} - {\textbf{Q}}_0)^2 + {\kappa_0}^2}
\label{eq:eqn1}
\end{equation}

where $I(\textbf{Q})$, which serves as the instrumental resolution function $R(\textbf{Q})$), is the measured intensity at the momentum transfer \textbf{Q}, $\textbf{Q}_0$ is the Bragg or superlattice peak position, $I_0$ is the integrated area of the peak and $\kappa_0$ the Half Width at Half Maximum (HWHM) of the resolution peak profile.

Above the transition temperature, the critical scattering data are fitted with the following functional form:

\begin{equation}
I(\textbf{Q}) = \int R({\textbf{Q}}^{'}) S(\textbf{Q}-{\textbf{Q}}^{'}) d{\textbf{Q}}^{'}
\label{eq:eqn2}
\end{equation}

where $I$(\textbf{Q}) is the measured peak intensity, $R$({\textbf{Q}}$^{'}$) the experimentally determined instrumental resolution (Eqn. (1)) and $S$(\textbf{Q}) another Lorentzian function, given by

\begin{equation}
S(\textbf{Q}) =  \frac{A}{(\textbf{Q} - {\textbf{Q}}_0)^2 + {\kappa}^2}
\label{eq:eqn3}
\end{equation}

where $A/{\kappa}^2$ and $\kappa$ are the staggered susceptibility and HWHM of the peak, respectively, in the critical scattering region above the transition temperature. $\kappa$= $1/\xi$, where $\xi$ is the correlation length.

\section{Crystal structure}

Select Ba4310 crystals were crushed and characterized via powder x-ray diffraction. Fig 1(d) displays the PXRD data collected at $T$ = 300K, which falls into the P$2_1$/$a$ space group (SG) (lattice parameters, $a$=5.784\AA$\,$, $b$=13.226\AA$\,$, $c$=7.239\AA$\,$ and $\beta$=112.935$^{\circ}$ at 300K)\cite{Wilkens1991}.

The basic building blocks of Ba4310 - Ir$_3$O$_{12}$ trimers - form a buckling quasi-two-dimensional (quasi-2D) layered structure (Fig. 1(a)-1(b)). Strong local distortions result in highly irregular Ir octahedra, largely unequal Ir-O bond lengths, and unequal O-Ir-O angles (Fig. 1(b)). Here, Ir1 denotes intra-trimer face-sharing Ir octahedra, and Ir2 inter-trimer corner-sharing Ir octahedra. The Ir2 octehedra have the most noticeable deviation from the ideal octahedron, where the bonds between the apical oxygen (O1 or O4) and Ir2 are up to 5\% elongated (2.124\AA$\,$) or compressed (1.92\AA$\,$) relative to the average bond length.

\section{Magnetometry and thermodynamic}

The magnetization data of the Ba4310 single crystals exhibit two transitions at 142.3(9)K and 24.8(5)K, respectively, consistent with observations in previously reported samples grown under a magnetic field (Fig. 2(a)) \cite{Cao2020_npjQM, cao2020quest}. Highly anisotropic magnetic responses are manifested when the external magnetic field is applied along different directions, with the largest response at fields perpendicular to the $ab$ plane, $i.e.$, the $c^*$ direction. The hard-axis is along the crystallographic $a$ direction. The large anisotropy in magnetization indicates the Ising nature of the Ir spin moments, which is not uncommon because of the low symmetry of the crystal structure.

The high-temperature transition, which is later verified to be a structural transition in origin, is also confirmed by a specific heat measurement after the phonon contribution has been subtracted, as shown in Fig. 2(b).

\section{x-ray scattering measurements}

In order to gain more insight about the structural and magnetic properties of Ba4310, REXS experiments on a Ba4310 single crystal were performed. The important experimental observation is the emerging superlattice peaks, which are summarized in Fig. 3(a)-3(b). Major structural (grey) and magnetic (green) ordering peaks at the base temperature are presented, where the marker sizes are proportional to the relative peak intensities.

\begin{figure}[t]
\centering
\includegraphics*[width=6.5cm]{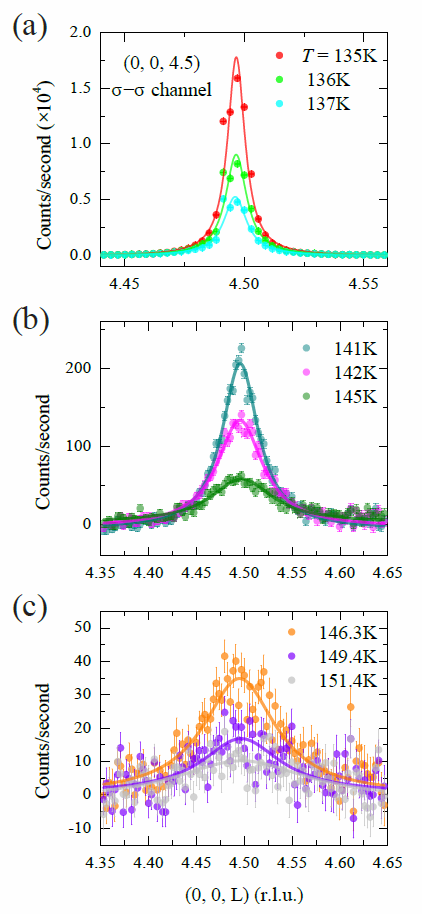}
\caption{(Color online) Temperature dependence of the structural superlattice peak (0, 0, 4.5), collected in $\sigma$-$\sigma$ channel at energy $E$ = 11.215 keV. (a)-(c), Background subtracted longitudinal $L$ scans of (0, 0, 4.5) peak at select temperatures across $T_{\text{S}}$. Solid lines are Lorentzian fits (Eqn.(\ref{eq:eqn1})- (\ref{eq:eqn2})) to the experimental data.}
\label{fig:Fig2_alpha}
\end{figure}

\begin{figure}[t]
\centering
\includegraphics*[width=6.2 cm]{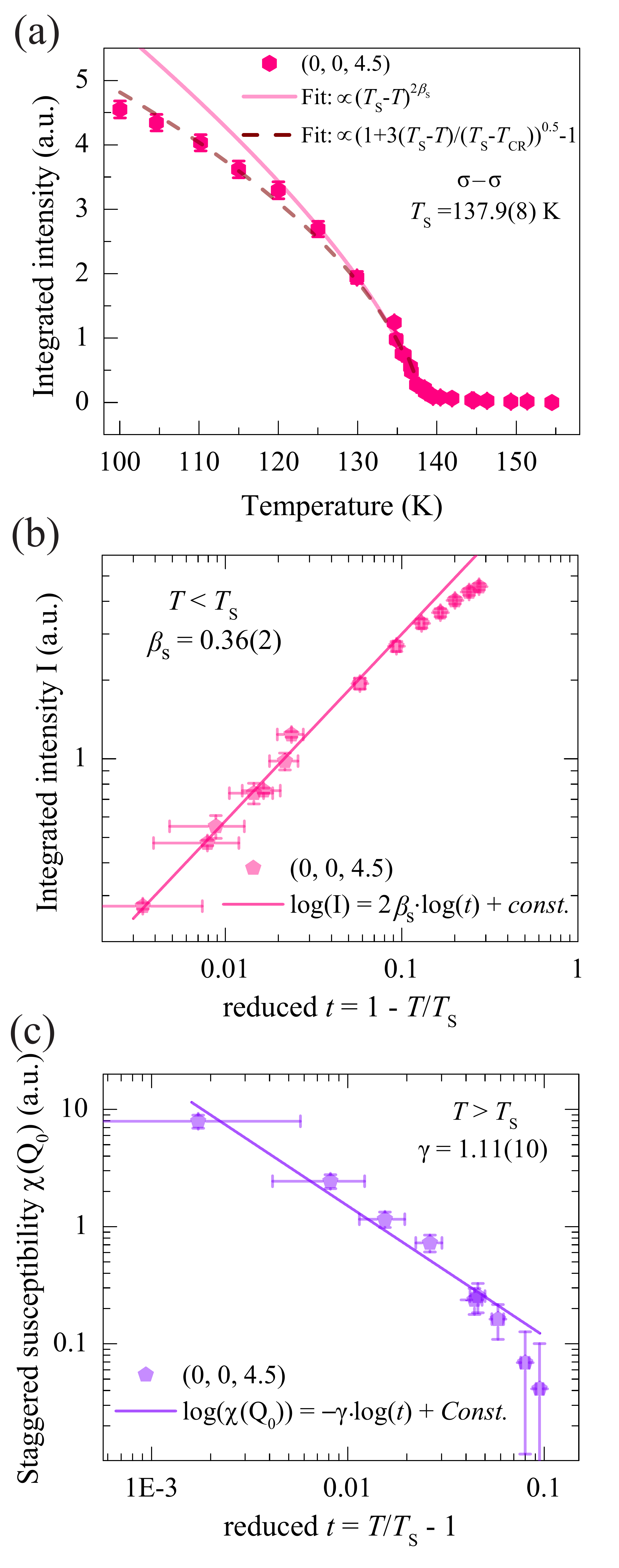}
\caption{(Color online) (a) Temperature dependence of integrated intensity of the structural superlattice peak (0, 0, 4.5), which is normalized by the intensity of reference nuclear Bragg (0, 0, 5) peak. Pink solid line is a power law fit to the intensity: $I\propto(T_{\text{S}} - T)^{2\beta_s} $. Red dash line is the tricritical-mean field cross over fit (Eqn. (\ref{eq:eqn5})). (b-c) log-log plot of the integrated intensity (b) or structural susceptibility $\chi_d$ (c) at $T$ $<$ $T_{\text{S}}$ or $T$ $>$ $T_{\text{S}}$, respectively. The reduced temperature is defined as $t$ = $|$1 - $T/T_{\text{S}}|$. Linear fits to data yield $\beta_{\text{s}} = 0.36(2)$ and $\gamma=1.11(10)$.}
\label{fig:Fig2_beta}
\end{figure}

\begin{figure}[t]
\centering
\includegraphics*[width=6.6cm]{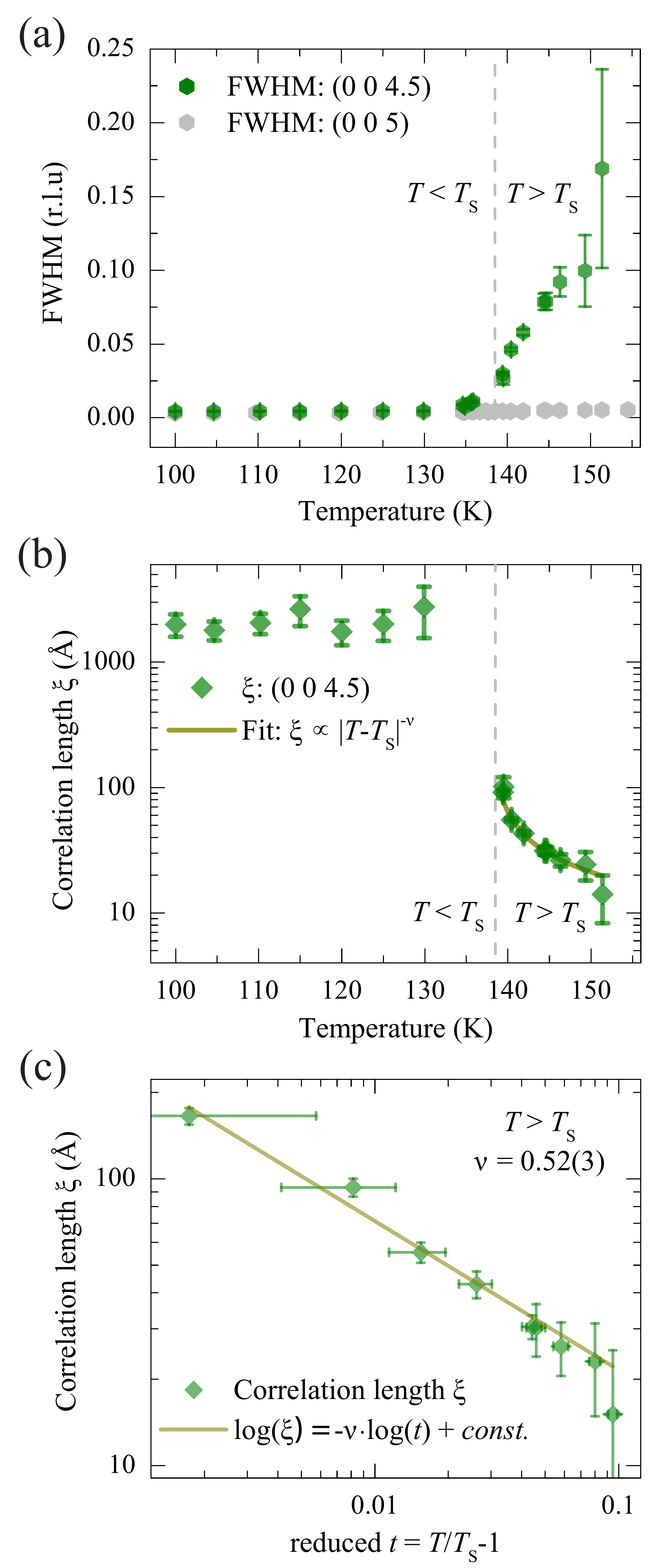}
\caption{(Color online) (a) Temperature dependence of the FWHM of (0 0 4.5) and (0 0 5) peaks. (b) Temperature dependence of structural correlation length $\xi$, deduced from the peak widths of (0 0 4.5) after deconvolving the instrument resolution (fit from the reference peak (0 0 5)). Solid yellow line is the power law fit ($\xi$ $\propto {\mid}T-T_{\text{S}}{\mid}^{-\nu}$) to correlation lengths above $T_{\text{S}}$. (c) Log-log plot of $\xi$ v.s. reduced temperature $t$ above $T_{\text{S}}$. A linear fit to data yields $\nu$ = 0.52(3). Vertical dash lines in (a)-(b) denote the structural transition temperature.}
\label{fig:Fig2_gamma}
\end{figure}

\begin{figure}[t]
\centering
\includegraphics*[width=6.5cm]{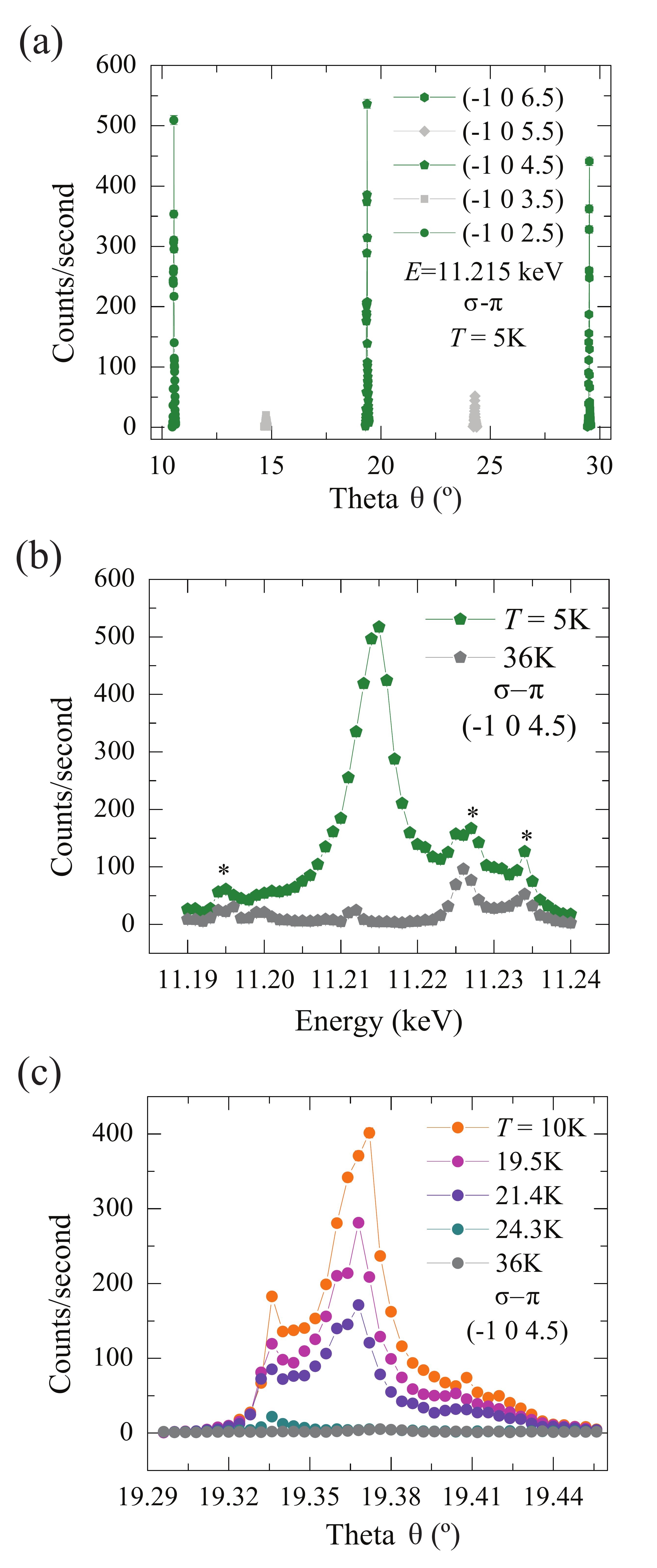}
\caption{(Color online) Magnetic superlattice peaks (-1 0 $L$+0.5) ($L$ = 2, 4, 6,...) in the $\sigma$-$\pi$ channel. (a) Theta scans of (-1 0 $L$+0.5) peaks, where $L$ are integers from 2 to 6. Green symbols: magnetic peaks and gray symbols: structural peaks. Data are collected at 11.215 keV in $\sigma$-$\pi$ channel at $T$ = 5K. (b) Energy dependence of (-1 0 4.5) peak at $T$ = 5K and $T$ = 36K, respectively. Peaks marked by asterisks indicate the temperature independent multiple scattering peaks. (c) Theta scans of (-1 0 4.5) peak at select temperatures.}
\label{fig:Fig3_alpha}
\end{figure}

\begin{figure}[t]
\centering
\includegraphics*[width=6.5cm]{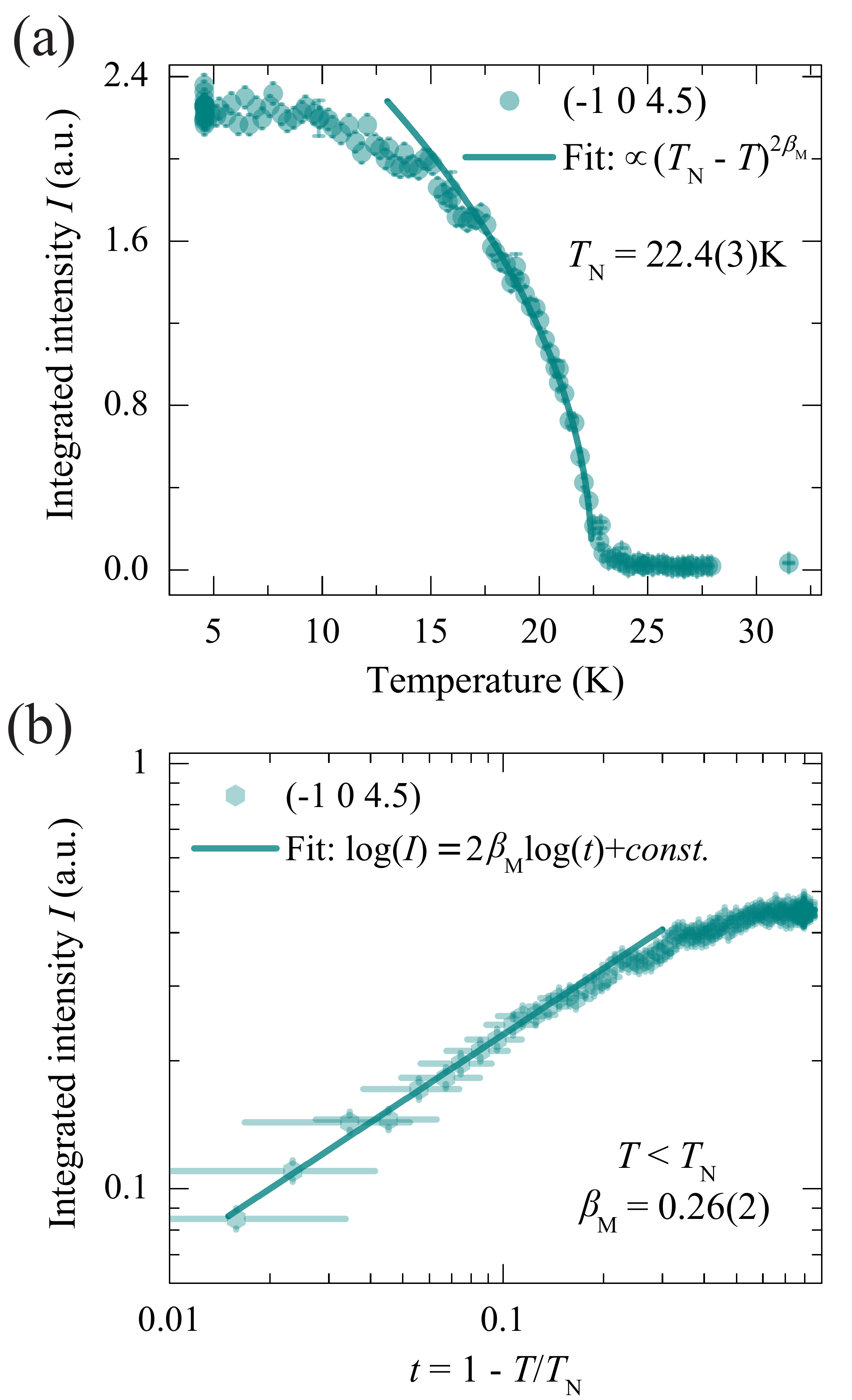}
\caption{(Color online) (a) Temperature dependence of the integrated intensity of magnetic (-1 0 4.5) peak in the $\sigma$-$\pi$ channel. Solid line is the power law fit to intensity: $I \propto (T_{\text{N}} - T)^{2\beta_M}$. (b) Log-log plot of data in (a) with reduced temperature $t$ = 1 - $T/T_{\text{N}}$. A linear fit is applied, resulting with $\beta_M$ = 0.26(2).}
\label{fig:Fig3_beta}
\end{figure}

\begin{figure}[t]
\centering
\includegraphics*[width=6.5cm]{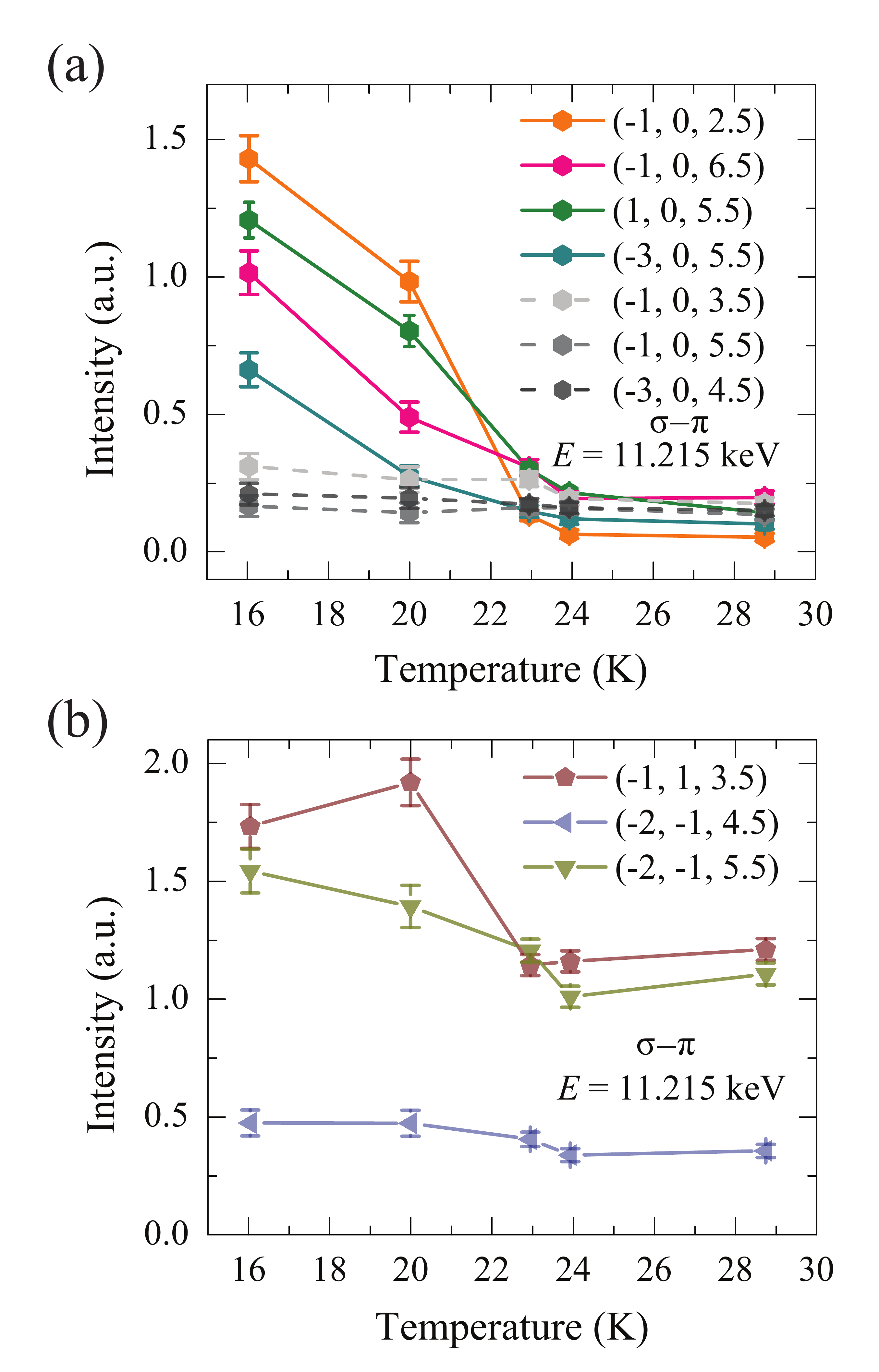}
\caption{(Color online)(a)-(b) Temperature dependence of the integrated intensity of additional select superlattice peaks, measured in $\sigma$-$\pi$ channel at 11.215 keV.}
\label{fig:Fig4}
\end{figure}

The most conspicuous $\textbf{Q}$ positions signaling the symmetry lowering at $T_{\text{S}}$ $\sim$ 142.3(9)K are the emerging superlattice peaks at (0 0 $L$+0.5) ($L$=integer values, Fig. 3(a), Fig. 4 and Fig. S1) \cite{supp}. Here, the peak intensity in the $\sigma$-$\sigma$ channel is much higher than in the $\sigma$-$\pi$ channel. Furthermore, energy scans show a dip at the Ir resonance energy (11.215~keV) (Fig. S1). Fig. 4 displays the temperature dependence of the longitudinal $L$ scans of (0 0 4.5) at 11.215 keV. A second order phase transition is clearly indicated from the power law analysis in Fig. 5(a)-5(b). To gain more insight about the nature of the structural transition, power law fitting of the order parameter \cite{Pelissetto2002,Birgeneau1977_PRB,Harris1995_PRB,Birgeneau_1999PRB}, the integrated area of the peak intensity, is applied near $T_{\text{S}}$ for reduced temperature $t$ = $|$1 - $T/T_{\text{S}}|$ $<$ 0.1 (pink solid lines in Fig. 5(a)-5(b)) under the form:

\begin{equation}
I \propto (T_{\text{S}} - T)^{2\beta_s}
\label{eq:eqn4}
\end{equation}

which yields fit parameters $T_{\text{S}}$ = 137.9(8)(K) and $\beta_s$ = 0.36(2). The slightly smaller $T_{\text{S}}$ obtained from the fit could be due to a beam heating effect.

The full width at half maximum (FWHM$=2\kappa_0$) of the structural superlattice (0 0 4.5) peak is resolution-limited below $\sim$130K (Eqn. (\ref{eq:eqn1}), Fig. 6(a)), indicating the long range order at low temperatures. Approaching $T_{\text{S}}$ from below (Fig. 6(a)), the peak starts broadening because of thermal fluctuations in addition to the Bragg scattering near the phase transition. Two separate functions are needed to adequately fit the peak when the temperature is approaching $T_{\text{S}}$ from below: a resolution limited Bragg peak on top of a broad fluctuating component. Above $T_{\text{S}}$, only the broad diffuse scattering arising from the dynamic fluctuations remains visible, and is well represented by a single Lorentian peak with broadened peak width (Eqn. (\ref{eq:eqn2})-(\ref{eq:eqn3}), Fig. 4(b)-4(c)). Fits to the broadened Lorentian, Eqn. (\ref{eq:eqn2})-(\ref{eq:eqn3}), yield the inverse correlation length $\xi$=$1/\kappa$ and the staggered susceptibility $S(\textbf{Q}_0)=A/\kappa^2$. Power law fitting of $\xi$ above $T_{\text{S}}$ with Eqn. (\ref{eq:eqn_xi}) yields $\nu$ = 0.52(3) (Fig. 6(b)-6(c)). The staggered susceptibility $\chi(\textbf{Q}_0) \sim S(\textbf{Q}_0)=A/\kappa^2$ also follows a power law Eqn. (\ref{eq:eqn_chi}). The results are shown in Fig. 5(c)), which yields $\gamma=1.11(10)$.

\begin{equation}
\xi \propto (T-T_{\text{S}})^{-\nu} 
\label{eq:eqn_xi}
\end{equation}

\begin{equation}
\chi(\textbf{Q}_0) \propto (T_{\text{S}}-T)^{-\gamma} 
\label{eq:eqn_chi}
\end{equation}

Clearly the staggered structural susceptibility and correlation length exhibit power laws ($\gamma=1.11(10)$ and $\nu=0.52(3)$, Figs. 5-6) consistent with mean field behavior ($\gamma=1$ and $\nu=0.5$), as commonly observed in second order structural phase transitions. However, the power law fit to the order parameter data, $\beta=0.36(2)$ (Fig. 5(a)), seems more consistent with critical behavior ($\beta$=0.32$\sim$0.37) rather than mean field behavior, $\beta=0.5$ \cite{Pelissetto2002, Birgeneau_1999PRB,Shashidhar1988_PRL}. This situation is commonly encountered in structural phase transitions in solids and in liquid crystals \cite{Shashidhar1988_PRL}. Most typically, structural phase transitions are first rather than second order; this means that in cases where a second order transition occurs there may be a nearby first order transition in the phase diagram. The point separating the second order line of transitions from the first order line is a tricritical point. At a tricritical point in 3 dimensions, mean field theory holds with $\gamma=1$, $\nu=0.5$ and $\beta=0.25$ \cite{Kim2002_PRL}. Thus, as has been extensively discussed elsewhere, mean field structural phase transitions in the vicinity of a tricritical point will exhibit a crossover in the temperature dependence of the order parameter from $\beta=0.25$ to $\beta=0.5$ power law behavior as the phase transition is approached and this may simulate an order parameter power law intermediate between $\beta=0.25$ and $\beta=0.5$. The explicit prediction for the order parameter $\phi$ in the vicinity of a tricritical point is \cite{Birgeneau_1999PRB}:

\begin{equation}
\phi = \phi_0((1+3(T_{\text{S}}-T)/(T_{\text{S}}-T_{\text{CR}}))^{0.5}-1)^{0.5}
\label{eq:eqn5}
\end{equation}

\begin{equation}
I \propto (1+3(T_{\text{S}}-T)/(T_{\text{S}}-T_{\text{CR}}))^{0.5}-1
\label{eq:eqn6}
\end{equation}

Here $T_{\text{CR}}$ is the adjustable tricritical to critical crossover temperature, $T_{\text{S}}$ the actual phase transition temperature and the exponent is fixed to be 0.5. This formula has the same number of adjustable parameters as the power law Eqn. (4). The results of fits to the mean field tricritical crossover model, Eqn. (\ref{eq:eqn6}), are shown as the dashed line in Fig. 5(a).  Clearly the mean field model works very well over an extended range of temperature. From this alternative model, one derives $T_{\text{S}}$ = 138.1(2)K and $T_{\text{CR}}$ = 135.5(8)K, implying the crossover from tricritical to mean field behavior occurs in two separate temperature scales. Thus, we conclude that Ba4310 shows an elegant example of a mean field, second order structural phase transition.

Further cooling below $T_{\text{N}}$ $\approx$ 25K, an additional set of magnetic superlattice peaks emerge at $L$+0.5 ($L$ = integers, Fig. 3(b), Fig. 7-9). Due to the monoclinic lattice structure, many structural peaks are suppressed in the $H0L$ plane. This provides a crucial basis for distinguishing the magnetic component from the otherwise much stronger structural/charge component of the half-$L$ super-modulation. For example, at $\textbf{Q}$=(-1 0 4.5), the photon energy dependence (Fig. 7(b)) and temperature dependence (Fig. 7(c) and Fig. 8) clearly demonstrate a second order phase transition at $T_N$ that is of magnetic nature, in addition to the structural transition at the higher $T_S$. A power law fit to the order parameter yields $T_{\text{N}}$ = 22.4(3)K and $\beta_M$ = 0.26(2) (Fig. 8). Again, a small offset in temperature is observed compared to $T_{\text{N}}$ $\approx$ 25K inferred from magnetization data. The $\beta_{M}$ value 0.26(2) is smaller than theoretical predictions from the expected 3D Ising behavior $\beta = 0.32$ \cite{Pelissetto2002}. Similar values of $\beta$ have been reported in the magnetic transitions of other layered iridates, including the 5$d$ cuprate analogues Ba$_2$IrO$_4$ ($\beta=0.25(3)$ \cite{Boseggia2013_PRL}) and Sr$_2$IrO$_4$ ($\beta=0.20(2)$ \cite{Dhital2013}). Considering the Ising nature of this quasi-2D magnet, a better description of the magnetic phase transition would be a gradual crossover from the initial 2D Ising-like ($\beta$=0.125) to the asymptotic 3D Ising ($\beta$=0.32) near the phase boundary. Therefore, an effective exponent $\beta_{\text{eff}}$ = 0.2$\sim$0.25 may be manifestly observed in Ba4310 and other quasi-2D Ising systems.

It is worth emphasizing that the structural and magnetic superlattice peaks can be comparable in intensity, and occupy the same $\textbf{Q}$ positions - such as at $\textbf{Q}$ = (-1 1 3.5) and (-2 -1 5.5) (Fig.~\ref{fig:Fig4}(b)). When the structural contributions dominate (as is the case in $0KL$ plane), no pronounced magnetic component is seen. In the $H0L$ plane, however, magnetic and structural superlattice reflections become distinguishable - at a given $\textbf{Q}$ = ($H$ 0 $L$+0.5), it is either predominantly structural or predominantly magnetic. Such a pattern is summarized in Fig. 3(b), where the magnetic and structural superlattice peaks interlace each other in the $H0L$ plane. Subsequently, strong constraints can be imposed on the spin configuration of the system's magnetic ground state.

\section{Discussion and conclusions}

In Ba4310, the emergence of two types of superlattice peaks, $i.e.$ structural and magnetic, at the same $\textbf{Q}$ positions at separate temperature scales is now firmly established. Their representative occurrences in the 0KL and H0L planes are summarized in Fig. 3. The complete determination of the spin configuration in the magnetic ground state remains challenging, in part due to the lower-symmetry crystal structure below $T_S$. On the other hand, strongly anisotropic $g$-values are expected from the large anisotropy in magnetization (Fig. 2(a)). This strong anisotropy can arise from the large IrO$_6$ octahedra distortions and low structural symmetry evidenced from PXRD (Fig. 1). One complication is the resulting non-negligible difference between the spin moment $\textbf{S}$ and the auxiliary vector $\textbf{N}$ directions probed by REXS experiments \cite{Kubota2015PRB, Chaloupka2016PRB}.

Assuming that the magnetic moments are carried by Ir$^{4+}$ ions with $J_{\text{eff}}$=1/2 moment \cite{LagunaMarco2010PRL, Ju2013PRB} and an acceptable difference between $\textbf{S}$ and $\textbf{N}$, we approximate the structure with the room temperature space group P$2_1$/$a$~\cite{Wilkens1991, Cao2020_npjQM}. An irreducible representation analysis \cite{Wills2000} with a propagation vector $k$ = (0 0 0.5) on space group P$2_1$/$a$, however, fails to reproduce the momentum and intensity pattern of the magnetic peaks. Instead, a magnetic model assuming that the Ir magnetic moments follow the local distortions, $i.e.$, along the largest Ir-O bond directions, better captures the observed pattern. This simple, approximate magnetic model also accounts for the anisotropic magnetization data (Fig. 2(a)). Furthermore, it opens up the possibility of a field-induced spin-flop transition \cite{Cao2020_npjQM}, where the intra-trimer Ir1 might become canted and have a net moment along the $c^*$ direction under external magnetic fields. Such an approximate magnetic model with spin moments tracking local structural distortions is not unique compared to other analogous systems with face-sharing octahedra trimer units and strong spin-orbit coupling. For example, in Ba$_4$Ru$_3$O$_{10}$, the Ru spin moments are also considered to follow the shortest Ru-O bond directions at both ends of the RuO$_6$-trimer \cite{Klein2011PRB}. We note that, a complete determination of the magnetic ground state spin configuration, however, would require further knowledge of the electronic wave-functions of Ir with accurate $g$-factors and the ground state crystal structure of Ba4310.

Our experimental findings about the symmetry reduction at $T_S$ have several implications. First, the high temperature structural transition is a mean field second order transition, as evidenced by our independent power law analyses of the temperature dependence of the order parameter, correlation length and structural susceptibility directly accessed via x-ray scattering (Figs. 5-6) \cite{Pelissetto2002}. This is further corroborated by the specific heat critical divergence (Fig. 2(b)), and the lack of thermal hysteresis around $T_S$ from both the scattering and magnetization measurements.

Second, the magnetization anomaly at $T_{\text{S}}$ (Fig. 2(a)) implies the relevance of magneto-elastic coupling (MEC), which may conceivably facilitate the symmetry lowering. One source of the MEC is the change in the ground state wave function due to the atomic displacements caused by the transition. In fact, it is the pertinence of MEC or pseudospin-lattice coupling that justifies isothermal metamagnetic behavior and predicts the structural transition in the prototypical spin orbit coupled Mott insulator Sr$_2$IrO$_4$ \cite{Liu2019, Porras2019PRB,Cao1998PRB, Chen2015PRB, Chen2018NatComm}. These physical ideas are also widely applicable to other 4$d$/5$d$ transition metal oxides with strong spin-orbit coupling, such as Sr$_3$NiIrO$_6$ \cite{ONeal2019_npjQM} and Ba$_2$CeIrO$_6$ \cite{Revelli2019PRB}. Another major outcome from MEC is the strong impact of exchange couplings between Ir magnetic moments which are further amplified through spin-orbit entanglement and local distortions. This might explain the exceptional sensitivity of the magnetic ground state in Ba4310 \cite{Cao2020_npjQM, cao2020quest}. 

Third, the consecutive symmetry lowering transitions at the same wave vector - albeit separated for more than 100~K in temperature - highlights the nontrivial interplay between the magnetic and structural degrees of freedom in the Ba4310 system. Recent studies show that the high temperature structural transition and the low temperature magnetic order can be simultaneously suppressed, accentuating their closely intertwined relation \cite{Cao2020_npjQM, cao2020quest}. On the one hand, if the structural transition were to be induced by magnetism, substantial magnetic fluctuations are required to survive up to 6 times the actual magnetic transition temperature $T_{\text{N}}$. On the other hand, it appears that despite the entropy release at $T_{\text{S}}$, residual instability remains at the same momentum until the lower temperature magnetic order triggers a second transition into the simultaneous ground states of both the lattice and magnetic moments.

In summary, our study firmly establishes two consecutive second order phase transitions in Ba4310 at $T_{\text{S}}$ $\approx$ 142K (structural) and $T_{\text{N}}$ $\approx$ 25K (magnetic) at the same ordering wave vector. The magneto-elastic coupling, inferred from the magnetization anomaly at and above $T_{\text{S}}$, might be partly responsible for the structural transition. Such coupling also manifests itself through the Ir-O bond length modulated magnetic moment orientation. Our study highlights the intertwined structural and magnetic degrees of freedoms in Ba4310, which might shed light on novel approaches to producing unconventional magnetic ground states.

\bigbreak

The authors wish to thank Dung-Hai Lee, Zhenglu Li and Edwin W. Huang for fruitful discussions. XC also thanks Robert Kealhofer and James Analytis for help with specific heat measurements. Work at Lawrence Berkeley National Laboratory was funded by the U.S. Department of Energy, Office of Science, Office of Basic Energy Sciences, Materials Sciences and Engineering Division under Contract No. DE-AC02-05-CH11231 within the Quantum Materials Program (KC2202). YH acknowledges support from the Miller Institute for Basic Research in Science. AF acknowledges support from the Alfred P. Sloan Fellowship in Physics. This research used resources of the Advanced Photon Source, a U.S. Department of Energy (DOE) Office of Science User Facility, operated for the DOE Office of Science by Argonne National Laboratory under Contract No. DE-AC02-06CH11357. Extraordinary facility operations were supported in part by the DOE Office of Science through the National Virtual Biotechnology Laboratory, a consortium of DOE national laboratories focused on the response to COVID-19, with funding provided by the Coronavirus CARES Act. This work is based upon research conducted at the Center for High Energy X-ray Sciences (CHEXS) which is supported by the National Science Foundation under award DMR-1829070.

\bibliography{Ba43110_REXS}

\begin{thebibliography}{51}%
\makeatletter
\providecommand \@ifxundefined [1]{%
 \@ifx{#1\undefined}
}%
\providecommand \@ifnum [1]{%
 \ifnum #1\expandafter \@firstoftwo
 \else \expandafter \@secondoftwo
 \fi
}%
\providecommand \@ifx [1]{%
 \ifx #1\expandafter \@firstoftwo
 \else \expandafter \@secondoftwo
 \fi
}%
\providecommand \natexlab [1]{#1}%
\providecommand \enquote  [1]{``#1''}%
\providecommand \bibnamefont  [1]{#1}%
\providecommand \bibfnamefont [1]{#1}%
\providecommand \citenamefont [1]{#1}%
\providecommand \href@noop [0]{\@secondoftwo}%
\providecommand \href [0]{\begingroup \@sanitize@url \@href}%
\providecommand \@href[1]{\@@startlink{#1}\@@href}%
\providecommand \@@href[1]{\endgroup#1\@@endlink}%
\providecommand \@sanitize@url [0]{\catcode `\\12\catcode `\$12\catcode
  `\&12\catcode `\#12\catcode `\^12\catcode `\_12\catcode `\%12\relax}%
\providecommand \@@startlink[1]{}%
\providecommand \@@endlink[0]{}%
\providecommand \url  [0]{\begingroup\@sanitize@url \@url }%
\providecommand \@url [1]{\endgroup\@href {#1}{\urlprefix }}%
\providecommand \urlprefix  [0]{URL }%
\providecommand \Eprint [0]{\href }%
\providecommand \doibase [0]{https://doi.org/}%
\providecommand \selectlanguage [0]{\@gobble}%
\providecommand \bibinfo  [0]{\@secondoftwo}%
\providecommand \bibfield  [0]{\@secondoftwo}%
\providecommand \translation [1]{[#1]}%
\providecommand \BibitemOpen [0]{}%
\providecommand \bibitemStop [0]{}%
\providecommand \bibitemNoStop [0]{.\EOS\space}%
\providecommand \EOS [0]{\spacefactor3000\relax}%
\providecommand \BibitemShut  [1]{\csname bibitem#1\endcsname}%
\let\auto@bib@innerbib\@empty
\bibitem [{\citenamefont {Witczak-Krempa}\ \emph {et~al.}(2014)\citenamefont
  {Witczak-Krempa}, \citenamefont {Chen}, \citenamefont {Kim},\ and\
  \citenamefont {Balents}}]{Witczak-Krempa2014}%
  \BibitemOpen
  \bibfield  {author} {\bibinfo {author} {\bibfnamefont {W.}~\bibnamefont
  {Witczak-Krempa}}, \bibinfo {author} {\bibfnamefont {G.}~\bibnamefont
  {Chen}}, \bibinfo {author} {\bibfnamefont {Y.~B.}\ \bibnamefont {Kim}},\ and\
  \bibinfo {author} {\bibfnamefont {L.}~\bibnamefont {Balents}},\ }\href
  {https://doi.org/10.1146/annurev-conmatphys-020911-125138} {\bibfield
  {journal} {\bibinfo  {journal} {Annual Review of Condensed Matter Physics}\
  }\textbf {\bibinfo {volume} {5}},\ \bibinfo {pages} {57} (\bibinfo {year}
  {2014})}\BibitemShut {NoStop}%
\bibitem [{\citenamefont {Rau}\ \emph {et~al.}(2016)\citenamefont {Rau},
  \citenamefont {Lee},\ and\ \citenamefont {Kee}}]{Rau2016}%
  \BibitemOpen
  \bibfield  {author} {\bibinfo {author} {\bibfnamefont {J.~G.}\ \bibnamefont
  {Rau}}, \bibinfo {author} {\bibfnamefont {E.~K.-H.}\ \bibnamefont {Lee}},\
  and\ \bibinfo {author} {\bibfnamefont {H.-Y.}\ \bibnamefont {Kee}},\
  }\href@noop {} {\bibfield  {journal} {\bibinfo  {journal} {Annual Review of
  Condensed Matter Physics}\ }\textbf {\bibinfo {volume} {7}},\ \bibinfo
  {pages} {195} (\bibinfo {year} {2016})}\BibitemShut {NoStop}%
\bibitem [{\citenamefont {Cao}\ and\ \citenamefont
  {Schlottmann}(2018)}]{Cao2018RPP}%
  \BibitemOpen
  \bibfield  {author} {\bibinfo {author} {\bibfnamefont {G.}~\bibnamefont
  {Cao}}\ and\ \bibinfo {author} {\bibfnamefont {P.}~\bibnamefont
  {Schlottmann}},\ }\href {https://doi.org/10.1088/1361-6633/aaa979} {\bibfield
   {journal} {\bibinfo  {journal} {Reports on Progress in Physics}\ }\textbf
  {\bibinfo {volume} {81}},\ \bibinfo {pages} {042502} (\bibinfo {year}
  {2018})}\BibitemShut {NoStop}%
\bibitem [{\citenamefont {Bertinshaw}\ \emph {et~al.}(2019)\citenamefont
  {Bertinshaw}, \citenamefont {Kim}, \citenamefont {Khaliullin},\ and\
  \citenamefont {Kim}}]{Bertinshaw2019ARCMP}%
  \BibitemOpen
  \bibfield  {author} {\bibinfo {author} {\bibfnamefont {J.}~\bibnamefont
  {Bertinshaw}}, \bibinfo {author} {\bibfnamefont {Y.}~\bibnamefont {Kim}},
  \bibinfo {author} {\bibfnamefont {G.}~\bibnamefont {Khaliullin}},\ and\
  \bibinfo {author} {\bibfnamefont {B.}~\bibnamefont {Kim}},\ }\href
  {https://doi.org/10.1146/annurev-conmatphys-031218-013113} {\bibfield
  {journal} {\bibinfo  {journal} {Annual Review of Condensed Matter Physics}\
  }\textbf {\bibinfo {volume} {10}},\ \bibinfo {pages} {315} (\bibinfo {year}
  {2019})}\BibitemShut {NoStop}%
\bibitem [{\citenamefont {Balents}(2010)}]{Balents2010}%
  \BibitemOpen
  \bibfield  {author} {\bibinfo {author} {\bibfnamefont {L.}~\bibnamefont
  {Balents}},\ }\href {https://doi.org/10.1038/nature08917} {\bibfield
  {journal} {\bibinfo  {journal} {Nature}\ }\textbf {\bibinfo {volume} {464}},\
  \bibinfo {pages} {199} (\bibinfo {year} {2010})}\BibitemShut {NoStop}%
\bibitem [{\citenamefont {Zhou}\ \emph {et~al.}(2017)\citenamefont {Zhou},
  \citenamefont {Kanoda},\ and\ \citenamefont {Ng}}]{Zhou2017RMP}%
  \BibitemOpen
  \bibfield  {author} {\bibinfo {author} {\bibfnamefont {Y.}~\bibnamefont
  {Zhou}}, \bibinfo {author} {\bibfnamefont {K.}~\bibnamefont {Kanoda}},\ and\
  \bibinfo {author} {\bibfnamefont {T.-K.}\ \bibnamefont {Ng}},\ }\href@noop {}
  {\bibfield  {journal} {\bibinfo  {journal} {Rev. Mod. Phys.}\ }\textbf
  {\bibinfo {volume} {89}},\ \bibinfo {pages} {025003} (\bibinfo {year}
  {2017})}\BibitemShut {NoStop}%
\bibitem [{\citenamefont {Wan}\ \emph {et~al.}(2011)\citenamefont {Wan},
  \citenamefont {Turner}, \citenamefont {Vishwanath},\ and\ \citenamefont
  {Savrasov}}]{Wan2011}%
  \BibitemOpen
  \bibfield  {author} {\bibinfo {author} {\bibfnamefont {X.}~\bibnamefont
  {Wan}}, \bibinfo {author} {\bibfnamefont {A.~M.}\ \bibnamefont {Turner}},
  \bibinfo {author} {\bibfnamefont {A.}~\bibnamefont {Vishwanath}},\ and\
  \bibinfo {author} {\bibfnamefont {S.~Y.}\ \bibnamefont {Savrasov}},\
  }\href@noop {} {\bibfield  {journal} {\bibinfo  {journal} {Phys. Rev. B}\
  }\textbf {\bibinfo {volume} {83}},\ \bibinfo {pages} {205101} (\bibinfo
  {year} {2011})}\BibitemShut {NoStop}%
\bibitem [{\citenamefont {Wang}\ and\ \citenamefont
  {Senthil}(2011)}]{Wang2011PRL}%
  \BibitemOpen
  \bibfield  {author} {\bibinfo {author} {\bibfnamefont {F.}~\bibnamefont
  {Wang}}\ and\ \bibinfo {author} {\bibfnamefont {T.}~\bibnamefont {Senthil}},\
  }\href@noop {} {\bibfield  {journal} {\bibinfo  {journal} {Phys. Rev. Lett.}\
  }\textbf {\bibinfo {volume} {106}},\ \bibinfo {pages} {136402} (\bibinfo
  {year} {2011})}\BibitemShut {NoStop}%
\bibitem [{\citenamefont {Kim}\ \emph {et~al.}(2015)\citenamefont {Kim},
  \citenamefont {Sung}, \citenamefont {Denlinger},\ and\ \citenamefont
  {Kim}}]{Kim2015NatPhy}%
  \BibitemOpen
  \bibfield  {author} {\bibinfo {author} {\bibfnamefont {Y.~K.}\ \bibnamefont
  {Kim}}, \bibinfo {author} {\bibfnamefont {N.~H.}\ \bibnamefont {Sung}},
  \bibinfo {author} {\bibfnamefont {J.~D.}\ \bibnamefont {Denlinger}},\ and\
  \bibinfo {author} {\bibfnamefont {B.~J.}\ \bibnamefont {Kim}},\ }\href
  {https://doi.org/10.1038/nphys3503} {\bibfield  {journal} {\bibinfo
  {journal} {Nature Physics}\ }\textbf {\bibinfo {volume} {12}},\ \bibinfo
  {pages} {37} (\bibinfo {year} {2015})},\ \Eprint
  {https://arxiv.org/abs/1506.06639} {1506.06639} \BibitemShut {NoStop}%
\bibitem [{\citenamefont {Yan}\ \emph {et~al.}(2015)\citenamefont {Yan},
  \citenamefont {Ren}, \citenamefont {Xu}, \citenamefont {Xie}, \citenamefont
  {Tao}, \citenamefont {Choi}, \citenamefont {Lee}, \citenamefont {Choi},
  \citenamefont {Zhang},\ and\ \citenamefont {Feng}}]{Yan2015PRX}%
  \BibitemOpen
  \bibfield  {author} {\bibinfo {author} {\bibfnamefont {Y.~J.}\ \bibnamefont
  {Yan}}, \bibinfo {author} {\bibfnamefont {M.~Q.}\ \bibnamefont {Ren}},
  \bibinfo {author} {\bibfnamefont {H.~C.}\ \bibnamefont {Xu}}, \bibinfo
  {author} {\bibfnamefont {B.~P.}\ \bibnamefont {Xie}}, \bibinfo {author}
  {\bibfnamefont {R.}~\bibnamefont {Tao}}, \bibinfo {author} {\bibfnamefont
  {H.~Y.}\ \bibnamefont {Choi}}, \bibinfo {author} {\bibfnamefont
  {N.}~\bibnamefont {Lee}}, \bibinfo {author} {\bibfnamefont {Y.~J.}\
  \bibnamefont {Choi}}, \bibinfo {author} {\bibfnamefont {T.}~\bibnamefont
  {Zhang}},\ and\ \bibinfo {author} {\bibfnamefont {D.~L.}\ \bibnamefont
  {Feng}},\ }\href {https://doi.org/10.1103/PhysRevX.5.041018} {\bibfield
  {journal} {\bibinfo  {journal} {Phys. Rev. X}\ }\textbf {\bibinfo {volume}
  {5}},\ \bibinfo {pages} {041018} (\bibinfo {year} {2015})}\BibitemShut
  {NoStop}%
\bibitem [{\citenamefont {Kim}\ \emph {et~al.}(2008)\citenamefont {Kim},
  \citenamefont {Jin}, \citenamefont {Moon}, \citenamefont {Kim}, \citenamefont
  {Park}, \citenamefont {Leem}, \citenamefont {Yu}, \citenamefont {Noh},
  \citenamefont {Kim}, \citenamefont {Oh}, \citenamefont {Park}, \citenamefont
  {Durairaj}, \citenamefont {Cao},\ and\ \citenamefont {Rotenberg}}]{Kim2008}%
  \BibitemOpen
  \bibfield  {author} {\bibinfo {author} {\bibfnamefont {B.~J.}\ \bibnamefont
  {Kim}}, \bibinfo {author} {\bibfnamefont {H.}~\bibnamefont {Jin}}, \bibinfo
  {author} {\bibfnamefont {S.~J.}\ \bibnamefont {Moon}}, \bibinfo {author}
  {\bibfnamefont {J.-Y.}\ \bibnamefont {Kim}}, \bibinfo {author} {\bibfnamefont
  {B.-G.}\ \bibnamefont {Park}}, \bibinfo {author} {\bibfnamefont {C.~S.}\
  \bibnamefont {Leem}}, \bibinfo {author} {\bibfnamefont {J.}~\bibnamefont
  {Yu}}, \bibinfo {author} {\bibfnamefont {T.~W.}\ \bibnamefont {Noh}},
  \bibinfo {author} {\bibfnamefont {C.}~\bibnamefont {Kim}}, \bibinfo {author}
  {\bibfnamefont {S.-J.}\ \bibnamefont {Oh}}, \bibinfo {author} {\bibfnamefont
  {J.-H.}\ \bibnamefont {Park}}, \bibinfo {author} {\bibfnamefont
  {V.}~\bibnamefont {Durairaj}}, \bibinfo {author} {\bibfnamefont
  {G.}~\bibnamefont {Cao}},\ and\ \bibinfo {author} {\bibfnamefont
  {E.}~\bibnamefont {Rotenberg}},\ }\href@noop {} {\bibfield  {journal}
  {\bibinfo  {journal} {Phys. Rev. Lett.}\ }\textbf {\bibinfo {volume} {101}},\
  \bibinfo {pages} {076402} (\bibinfo {year} {2008})}\BibitemShut {NoStop}%
\bibitem [{\citenamefont {Kim}\ \emph {et~al.}(2009)\citenamefont {Kim},
  \citenamefont {Ohsumi}, \citenamefont {Komesu}, \citenamefont {Sakai},
  \citenamefont {Morita}, \citenamefont {Takagi},\ and\ \citenamefont
  {Arima}}]{Kim2009}%
  \BibitemOpen
  \bibfield  {author} {\bibinfo {author} {\bibfnamefont {B.~J.}\ \bibnamefont
  {Kim}}, \bibinfo {author} {\bibfnamefont {H.}~\bibnamefont {Ohsumi}},
  \bibinfo {author} {\bibfnamefont {T.}~\bibnamefont {Komesu}}, \bibinfo
  {author} {\bibfnamefont {S.}~\bibnamefont {Sakai}}, \bibinfo {author}
  {\bibfnamefont {T.}~\bibnamefont {Morita}}, \bibinfo {author} {\bibfnamefont
  {H.}~\bibnamefont {Takagi}},\ and\ \bibinfo {author} {\bibfnamefont
  {T.}~\bibnamefont {Arima}},\ }\href {https://doi.org/10.1126/science.1167106}
  {\bibfield  {journal} {\bibinfo  {journal} {Science}\ }\textbf {\bibinfo
  {volume} {323}},\ \bibinfo {pages} {1329} (\bibinfo {year}
  {2009})}\BibitemShut {NoStop}%
\bibitem [{\citenamefont {Jackeli}\ and\ \citenamefont
  {Khaliullin}(2009)}]{Jackeli2009PRL}%
  \BibitemOpen
  \bibfield  {author} {\bibinfo {author} {\bibfnamefont {G.}~\bibnamefont
  {Jackeli}}\ and\ \bibinfo {author} {\bibfnamefont {G.}~\bibnamefont
  {Khaliullin}},\ }\href {https://doi.org/10.1103/PhysRevLett.102.017205}
  {\bibfield  {journal} {\bibinfo  {journal} {Phys. Rev. Lett.}\ }\textbf
  {\bibinfo {volume} {102}},\ \bibinfo {pages} {017205} (\bibinfo {year}
  {2009})}\BibitemShut {NoStop}%
\bibitem [{\citenamefont {Kimchi}\ and\ \citenamefont
  {Vishwanath}(2014)}]{Kimchi2014}%
  \BibitemOpen
  \bibfield  {author} {\bibinfo {author} {\bibfnamefont {I.}~\bibnamefont
  {Kimchi}}\ and\ \bibinfo {author} {\bibfnamefont {A.}~\bibnamefont
  {Vishwanath}},\ }\href@noop {} {\bibfield  {journal} {\bibinfo  {journal}
  {Phys. Rev. B}\ }\textbf {\bibinfo {volume} {89}},\ \bibinfo {pages} {014414}
  (\bibinfo {year} {2014})}\BibitemShut {NoStop}%
\bibitem [{\citenamefont {Cao}\ \emph {et~al.}(2000)\citenamefont {Cao},
  \citenamefont {Crow}, \citenamefont {Guertin}, \citenamefont {Henning},
  \citenamefont {Homes}, \citenamefont {Strongin}, \citenamefont {Basov},\ and\
  \citenamefont {Lochner}}]{Cao_2000SSC}%
  \BibitemOpen
  \bibfield  {author} {\bibinfo {author} {\bibfnamefont {G.}~\bibnamefont
  {Cao}}, \bibinfo {author} {\bibfnamefont {J.}~\bibnamefont {Crow}}, \bibinfo
  {author} {\bibfnamefont {R.}~\bibnamefont {Guertin}}, \bibinfo {author}
  {\bibfnamefont {P.}~\bibnamefont {Henning}}, \bibinfo {author} {\bibfnamefont
  {C.}~\bibnamefont {Homes}}, \bibinfo {author} {\bibfnamefont
  {M.}~\bibnamefont {Strongin}}, \bibinfo {author} {\bibfnamefont
  {D.}~\bibnamefont {Basov}},\ and\ \bibinfo {author} {\bibfnamefont
  {E.}~\bibnamefont {Lochner}},\ }\href@noop {} {\bibfield  {journal} {\bibinfo
   {journal} {Solid State Communications}\ }\textbf {\bibinfo {volume} {113}},\
  \bibinfo {pages} {657 } (\bibinfo {year} {2000})}\BibitemShut {NoStop}%
\bibitem [{\citenamefont {Nguyen}\ and\ \citenamefont
  {Cava}(2019)}]{Nguyen_2019PRM}%
  \BibitemOpen
  \bibfield  {author} {\bibinfo {author} {\bibfnamefont {L.~T.}\ \bibnamefont
  {Nguyen}}\ and\ \bibinfo {author} {\bibfnamefont {R.~J.}\ \bibnamefont
  {Cava}},\ }\href {https://doi.org/10.1103/PhysRevMaterials.3.014412}
  {\bibfield  {journal} {\bibinfo  {journal} {Phys. Rev. Materials}\ }\textbf
  {\bibinfo {volume} {3}},\ \bibinfo {pages} {014412} (\bibinfo {year}
  {2019})}\BibitemShut {NoStop}%
\bibitem [{\citenamefont {Cao}\ \emph {et~al.}(2020{\natexlab{a}})\citenamefont
  {Cao}, \citenamefont {Zheng}, \citenamefont {Zhao}, \citenamefont {Ni},
  \citenamefont {Pocs}, \citenamefont {Zhang}, \citenamefont {Ye},
  \citenamefont {Hoffmann}, \citenamefont {Wang}, \citenamefont {Lee},
  \citenamefont {Hermele},\ and\ \citenamefont {Kimchi}}]{Cao2020_npjQM}%
  \BibitemOpen
  \bibfield  {author} {\bibinfo {author} {\bibfnamefont {G.}~\bibnamefont
  {Cao}}, \bibinfo {author} {\bibfnamefont {H.}~\bibnamefont {Zheng}}, \bibinfo
  {author} {\bibfnamefont {H.}~\bibnamefont {Zhao}}, \bibinfo {author}
  {\bibfnamefont {Y.}~\bibnamefont {Ni}}, \bibinfo {author} {\bibfnamefont
  {C.~A.}\ \bibnamefont {Pocs}}, \bibinfo {author} {\bibfnamefont
  {Y.}~\bibnamefont {Zhang}}, \bibinfo {author} {\bibfnamefont
  {F.}~\bibnamefont {Ye}}, \bibinfo {author} {\bibfnamefont {C.}~\bibnamefont
  {Hoffmann}}, \bibinfo {author} {\bibfnamefont {X.}~\bibnamefont {Wang}},
  \bibinfo {author} {\bibfnamefont {M.}~\bibnamefont {Lee}}, \bibinfo {author}
  {\bibfnamefont {M.}~\bibnamefont {Hermele}},\ and\ \bibinfo {author}
  {\bibfnamefont {I.}~\bibnamefont {Kimchi}},\ }\href@noop {} {\bibfield
  {journal} {\bibinfo  {journal} {npj Quantum Materials}\ }\textbf {\bibinfo
  {volume} {5}},\ \bibinfo {pages} {26} (\bibinfo {year}
  {2020}{\natexlab{a}})}\BibitemShut {NoStop}%
\bibitem [{\citenamefont {Laguna-Marco}\ \emph {et~al.}(2010)\citenamefont
  {Laguna-Marco}, \citenamefont {Haskel}, \citenamefont {Souza-Neto},
  \citenamefont {Lang}, \citenamefont {Krishnamurthy}, \citenamefont {Chikara},
  \citenamefont {Cao},\ and\ \citenamefont {van
  Veenendaal}}]{LagunaMarco2010PRL}%
  \BibitemOpen
  \bibfield  {author} {\bibinfo {author} {\bibfnamefont {M.~A.}\ \bibnamefont
  {Laguna-Marco}}, \bibinfo {author} {\bibfnamefont {D.}~\bibnamefont
  {Haskel}}, \bibinfo {author} {\bibfnamefont {N.}~\bibnamefont {Souza-Neto}},
  \bibinfo {author} {\bibfnamefont {J.~C.}\ \bibnamefont {Lang}}, \bibinfo
  {author} {\bibfnamefont {V.~V.}\ \bibnamefont {Krishnamurthy}}, \bibinfo
  {author} {\bibfnamefont {S.}~\bibnamefont {Chikara}}, \bibinfo {author}
  {\bibfnamefont {G.}~\bibnamefont {Cao}},\ and\ \bibinfo {author}
  {\bibfnamefont {M.}~\bibnamefont {van Veenendaal}},\ }\href
  {https://doi.org/10.1103/PhysRevLett.105.216407} {\bibfield  {journal}
  {\bibinfo  {journal} {Phys. Rev. Lett.}\ }\textbf {\bibinfo {volume} {105}},\
  \bibinfo {pages} {216407} (\bibinfo {year} {2010})}\BibitemShut {NoStop}%
\bibitem [{\citenamefont {Okazaki}\ \emph {et~al.}(2018)\citenamefont
  {Okazaki}, \citenamefont {Ito}, \citenamefont {Tanabe}, \citenamefont
  {Taniguchi}, \citenamefont {Ikemoto}, \citenamefont {Moriwaki},\ and\
  \citenamefont {Terasaki}}]{Okazaki_2018PRB}%
  \BibitemOpen
  \bibfield  {author} {\bibinfo {author} {\bibfnamefont {R.}~\bibnamefont
  {Okazaki}}, \bibinfo {author} {\bibfnamefont {S.}~\bibnamefont {Ito}},
  \bibinfo {author} {\bibfnamefont {K.}~\bibnamefont {Tanabe}}, \bibinfo
  {author} {\bibfnamefont {H.}~\bibnamefont {Taniguchi}}, \bibinfo {author}
  {\bibfnamefont {Y.}~\bibnamefont {Ikemoto}}, \bibinfo {author} {\bibfnamefont
  {T.}~\bibnamefont {Moriwaki}},\ and\ \bibinfo {author} {\bibfnamefont
  {I.}~\bibnamefont {Terasaki}},\ }\href
  {https://doi.org/10.1103/PhysRevB.98.205131} {\bibfield  {journal} {\bibinfo
  {journal} {Phys. Rev. B}\ }\textbf {\bibinfo {volume} {98}},\ \bibinfo
  {pages} {205131} (\bibinfo {year} {2018})}\BibitemShut {NoStop}%
\bibitem [{\citenamefont {Xu}\ \emph {et~al.}(2019)\citenamefont {Xu},
  \citenamefont {Yadav}, \citenamefont {Yushankhai}, \citenamefont
  {Siurakshina}, \citenamefont {van~den Brink},\ and\ \citenamefont
  {Hozoi}}]{Xu_2019PRB}%
  \BibitemOpen
  \bibfield  {author} {\bibinfo {author} {\bibfnamefont {L.}~\bibnamefont
  {Xu}}, \bibinfo {author} {\bibfnamefont {R.}~\bibnamefont {Yadav}}, \bibinfo
  {author} {\bibfnamefont {V.}~\bibnamefont {Yushankhai}}, \bibinfo {author}
  {\bibfnamefont {L.}~\bibnamefont {Siurakshina}}, \bibinfo {author}
  {\bibfnamefont {J.}~\bibnamefont {van~den Brink}},\ and\ \bibinfo {author}
  {\bibfnamefont {L.}~\bibnamefont {Hozoi}},\ }\href
  {https://doi.org/10.1103/PhysRevB.99.115119} {\bibfield  {journal} {\bibinfo
  {journal} {Phys. Rev. B}\ }\textbf {\bibinfo {volume} {99}},\ \bibinfo
  {pages} {115119} (\bibinfo {year} {2019})}\BibitemShut {NoStop}%
\bibitem [{\citenamefont {Kugel}\ \emph {et~al.}(2015)\citenamefont {Kugel},
  \citenamefont {Khomskii}, \citenamefont {Sboychakov},\ and\ \citenamefont
  {Streltsov}}]{Kugel2015}%
  \BibitemOpen
  \bibfield  {author} {\bibinfo {author} {\bibfnamefont {K.~I.}\ \bibnamefont
  {Kugel}}, \bibinfo {author} {\bibfnamefont {D.~I.}\ \bibnamefont {Khomskii}},
  \bibinfo {author} {\bibfnamefont {A.~O.}\ \bibnamefont {Sboychakov}},\ and\
  \bibinfo {author} {\bibfnamefont {S.~V.}\ \bibnamefont {Streltsov}},\
  }\href@noop {} {\bibfield  {journal} {\bibinfo  {journal} {Phys. Rev. B}\
  }\textbf {\bibinfo {volume} {91}},\ \bibinfo {pages} {155125} (\bibinfo
  {year} {2015})}\BibitemShut {NoStop}%
\bibitem [{\citenamefont {Cao}\ \emph {et~al.}(2020{\natexlab{b}})\citenamefont
  {Cao}, \citenamefont {Zhao}, \citenamefont {Bing}, \citenamefont {Pellatz},
  \citenamefont {Reznik}, \citenamefont {Schlottmann},\ and\ \citenamefont
  {Kimchi}}]{cao2020quest}%
  \BibitemOpen
  \bibfield  {author} {\bibinfo {author} {\bibfnamefont {G.}~\bibnamefont
  {Cao}}, \bibinfo {author} {\bibfnamefont {H.}~\bibnamefont {Zhao}}, \bibinfo
  {author} {\bibfnamefont {H.}~\bibnamefont {Bing}}, \bibinfo {author}
  {\bibfnamefont {N.}~\bibnamefont {Pellatz}}, \bibinfo {author} {\bibfnamefont
  {D.}~\bibnamefont {Reznik}}, \bibinfo {author} {\bibfnamefont
  {P.}~\bibnamefont {Schlottmann}},\ and\ \bibinfo {author} {\bibfnamefont
  {I.}~\bibnamefont {Kimchi}},\ }\href
  {https://doi.org/10.1038/s41535-020-00286-2} {\bibfield  {journal} {\bibinfo
  {journal} {npj Quantum Materials}\ }\textbf {\bibinfo {volume} {5}},\
  \bibinfo {pages} {83} (\bibinfo {year} {2020}{\natexlab{b}})}\BibitemShut
  {NoStop}%
\bibitem [{\citenamefont {Blume}(1985)}]{Blume1985}%
  \BibitemOpen
  \bibfield  {author} {\bibinfo {author} {\bibfnamefont {M.}~\bibnamefont
  {Blume}},\ }\href@noop {} {\bibfield  {journal} {\bibinfo  {journal} {Journal
  of Applied Physics}\ }\textbf {\bibinfo {volume} {57}},\ \bibinfo {pages}
  {3615} (\bibinfo {year} {1985})}\BibitemShut {NoStop}%
\bibitem [{\citenamefont {Hannon}\ \emph {et~al.}(1988)\citenamefont {Hannon},
  \citenamefont {Trammell}, \citenamefont {Blume},\ and\ \citenamefont
  {Gibbs}}]{Hannon1988PRL}%
  \BibitemOpen
  \bibfield  {author} {\bibinfo {author} {\bibfnamefont {J.~P.}\ \bibnamefont
  {Hannon}}, \bibinfo {author} {\bibfnamefont {G.~T.}\ \bibnamefont
  {Trammell}}, \bibinfo {author} {\bibfnamefont {M.}~\bibnamefont {Blume}},\
  and\ \bibinfo {author} {\bibfnamefont {D.}~\bibnamefont {Gibbs}},\ }\href
  {https://doi.org/10.1103/PhysRevLett.61.1245} {\bibfield  {journal} {\bibinfo
   {journal} {Phys. Rev. Lett.}\ }\textbf {\bibinfo {volume} {61}},\ \bibinfo
  {pages} {1245} (\bibinfo {year} {1988})}\BibitemShut {NoStop}%
\bibitem [{\citenamefont {Chen}\ \emph {et~al.}(2015)\citenamefont {Chen},
  \citenamefont {Hogan}, \citenamefont {Walkup}, \citenamefont {Zhou},
  \citenamefont {Pokharel}, \citenamefont {Yao}, \citenamefont {Tian},
  \citenamefont {Ward}, \citenamefont {Zhao}, \citenamefont {Parshall},
  \citenamefont {Opeil}, \citenamefont {Lynn}, \citenamefont {Madhavan},\ and\
  \citenamefont {Wilson}}]{Chen2015PRB}%
  \BibitemOpen
  \bibfield  {author} {\bibinfo {author} {\bibfnamefont {X.}~\bibnamefont
  {Chen}}, \bibinfo {author} {\bibfnamefont {T.}~\bibnamefont {Hogan}},
  \bibinfo {author} {\bibfnamefont {D.}~\bibnamefont {Walkup}}, \bibinfo
  {author} {\bibfnamefont {W.}~\bibnamefont {Zhou}}, \bibinfo {author}
  {\bibfnamefont {M.}~\bibnamefont {Pokharel}}, \bibinfo {author}
  {\bibfnamefont {M.}~\bibnamefont {Yao}}, \bibinfo {author} {\bibfnamefont
  {W.}~\bibnamefont {Tian}}, \bibinfo {author} {\bibfnamefont {T.~Z.}\
  \bibnamefont {Ward}}, \bibinfo {author} {\bibfnamefont {Y.}~\bibnamefont
  {Zhao}}, \bibinfo {author} {\bibfnamefont {D.}~\bibnamefont {Parshall}},
  \bibinfo {author} {\bibfnamefont {C.}~\bibnamefont {Opeil}}, \bibinfo
  {author} {\bibfnamefont {J.~W.}\ \bibnamefont {Lynn}}, \bibinfo {author}
  {\bibfnamefont {V.}~\bibnamefont {Madhavan}},\ and\ \bibinfo {author}
  {\bibfnamefont {S.~D.}\ \bibnamefont {Wilson}},\ }\href
  {https://doi.org/10.1103/PhysRevB.92.075125} {\bibfield  {journal} {\bibinfo
  {journal} {Phys. Rev. B}\ }\textbf {\bibinfo {volume} {92}},\ \bibinfo
  {pages} {075125} (\bibinfo {year} {2015})}\BibitemShut {NoStop}%
\bibitem [{\citenamefont {Fujiyama}\ \emph {et~al.}(2012)\citenamefont
  {Fujiyama}, \citenamefont {Ohsumi}, \citenamefont {Komesu}, \citenamefont
  {Matsuno}, \citenamefont {Kim}, \citenamefont {Takata}, \citenamefont
  {Arima},\ and\ \citenamefont {Takagi}}]{Fujiyama2012PRL}%
  \BibitemOpen
  \bibfield  {author} {\bibinfo {author} {\bibfnamefont {S.}~\bibnamefont
  {Fujiyama}}, \bibinfo {author} {\bibfnamefont {H.}~\bibnamefont {Ohsumi}},
  \bibinfo {author} {\bibfnamefont {T.}~\bibnamefont {Komesu}}, \bibinfo
  {author} {\bibfnamefont {J.}~\bibnamefont {Matsuno}}, \bibinfo {author}
  {\bibfnamefont {B.~J.}\ \bibnamefont {Kim}}, \bibinfo {author} {\bibfnamefont
  {M.}~\bibnamefont {Takata}}, \bibinfo {author} {\bibfnamefont
  {T.}~\bibnamefont {Arima}},\ and\ \bibinfo {author} {\bibfnamefont
  {H.}~\bibnamefont {Takagi}},\ }\href@noop {} {\bibfield  {journal} {\bibinfo
  {journal} {Phys. Rev. Lett.}\ }\textbf {\bibinfo {volume} {108}},\ \bibinfo
  {pages} {247212} (\bibinfo {year} {2012})}\BibitemShut {NoStop}%
\bibitem [{\citenamefont {Boseggia}\ \emph
  {et~al.}(2013{\natexlab{a}})\citenamefont {Boseggia}, \citenamefont {Walker},
  \citenamefont {Vale}, \citenamefont {Springell}, \citenamefont {Feng},
  \citenamefont {Perry}, \citenamefont {Sala}, \citenamefont {R{\o}nnow},
  \citenamefont {Collins},\ and\ \citenamefont {McMorrow}}]{Boseggia2013JPCM}%
  \BibitemOpen
  \bibfield  {author} {\bibinfo {author} {\bibfnamefont {S.}~\bibnamefont
  {Boseggia}}, \bibinfo {author} {\bibfnamefont {H.~C.}\ \bibnamefont
  {Walker}}, \bibinfo {author} {\bibfnamefont {J.}~\bibnamefont {Vale}},
  \bibinfo {author} {\bibfnamefont {R.}~\bibnamefont {Springell}}, \bibinfo
  {author} {\bibfnamefont {Z.}~\bibnamefont {Feng}}, \bibinfo {author}
  {\bibfnamefont {R.~S.}\ \bibnamefont {Perry}}, \bibinfo {author}
  {\bibfnamefont {M.~M.}\ \bibnamefont {Sala}}, \bibinfo {author}
  {\bibfnamefont {H.~M.}\ \bibnamefont {R{\o}nnow}}, \bibinfo {author}
  {\bibfnamefont {S.~P.}\ \bibnamefont {Collins}},\ and\ \bibinfo {author}
  {\bibfnamefont {D.~F.}\ \bibnamefont {McMorrow}},\ }\href
  {https://doi.org/10.1088/0953-8984/25/42/422202} {\bibfield  {journal}
  {\bibinfo  {journal} {Journal of Physics: Condensed Matter}\ }\textbf
  {\bibinfo {volume} {25}},\ \bibinfo {pages} {422202} (\bibinfo {year}
  {2013}{\natexlab{a}})}\BibitemShut {NoStop}%
\bibitem [{\citenamefont {Clancy}\ \emph {et~al.}(2014)\citenamefont {Clancy},
  \citenamefont {Lupascu}, \citenamefont {Gretarsson}, \citenamefont {Islam},
  \citenamefont {Hu}, \citenamefont {Casa}, \citenamefont {Nelson},
  \citenamefont {LaMarra}, \citenamefont {Cao},\ and\ \citenamefont
  {Kim}}]{Clancy2014PRB}%
  \BibitemOpen
  \bibfield  {author} {\bibinfo {author} {\bibfnamefont {J.~P.}\ \bibnamefont
  {Clancy}}, \bibinfo {author} {\bibfnamefont {A.}~\bibnamefont {Lupascu}},
  \bibinfo {author} {\bibfnamefont {H.}~\bibnamefont {Gretarsson}}, \bibinfo
  {author} {\bibfnamefont {Z.}~\bibnamefont {Islam}}, \bibinfo {author}
  {\bibfnamefont {Y.~F.}\ \bibnamefont {Hu}}, \bibinfo {author} {\bibfnamefont
  {D.}~\bibnamefont {Casa}}, \bibinfo {author} {\bibfnamefont {C.~S.}\
  \bibnamefont {Nelson}}, \bibinfo {author} {\bibfnamefont {S.~C.}\
  \bibnamefont {LaMarra}}, \bibinfo {author} {\bibfnamefont {G.}~\bibnamefont
  {Cao}},\ and\ \bibinfo {author} {\bibfnamefont {Y.-J.}\ \bibnamefont {Kim}},\
  }\href {https://doi.org/10.1103/PhysRevB.89.054409} {\bibfield  {journal}
  {\bibinfo  {journal} {Phys. Rev. B}\ }\textbf {\bibinfo {volume} {89}},\
  \bibinfo {pages} {054409} (\bibinfo {year} {2014})}\BibitemShut {NoStop}%
\bibitem [{\citenamefont {{Hwan Chun}}\ \emph {et~al.}(2015)\citenamefont
  {{Hwan Chun}}, \citenamefont {Kim}, \citenamefont {Kim}, \citenamefont
  {Zheng}, \citenamefont {Stoumpos}, \citenamefont {Malliakas}, \citenamefont
  {Mitchell}, \citenamefont {Mehlawat}, \citenamefont {Singh}, \citenamefont
  {Choi}, \citenamefont {Gog}, \citenamefont {Al-Zein}, \citenamefont {Sala},
  \citenamefont {Krisch}, \citenamefont {Chaloupka}, \citenamefont {Jackeli},
  \citenamefont {Khaliullin},\ and\ \citenamefont {Kim}}]{HwanChun2015NatPhy}%
  \BibitemOpen
  \bibfield  {author} {\bibinfo {author} {\bibfnamefont {S.}~\bibnamefont
  {{Hwan Chun}}}, \bibinfo {author} {\bibfnamefont {J.-W.}\ \bibnamefont
  {Kim}}, \bibinfo {author} {\bibfnamefont {J.}~\bibnamefont {Kim}}, \bibinfo
  {author} {\bibfnamefont {H.}~\bibnamefont {Zheng}}, \bibinfo {author}
  {\bibfnamefont {C.~C.}\ \bibnamefont {Stoumpos}}, \bibinfo {author}
  {\bibfnamefont {C.~D.}\ \bibnamefont {Malliakas}}, \bibinfo {author}
  {\bibfnamefont {J.~F.}\ \bibnamefont {Mitchell}}, \bibinfo {author}
  {\bibfnamefont {K.}~\bibnamefont {Mehlawat}}, \bibinfo {author}
  {\bibfnamefont {Y.}~\bibnamefont {Singh}}, \bibinfo {author} {\bibfnamefont
  {Y.}~\bibnamefont {Choi}}, \bibinfo {author} {\bibfnamefont {T.}~\bibnamefont
  {Gog}}, \bibinfo {author} {\bibfnamefont {A.}~\bibnamefont {Al-Zein}},
  \bibinfo {author} {\bibfnamefont {M.~M.}\ \bibnamefont {Sala}}, \bibinfo
  {author} {\bibfnamefont {M.}~\bibnamefont {Krisch}}, \bibinfo {author}
  {\bibfnamefont {J.}~\bibnamefont {Chaloupka}}, \bibinfo {author}
  {\bibfnamefont {G.}~\bibnamefont {Jackeli}}, \bibinfo {author} {\bibfnamefont
  {G.}~\bibnamefont {Khaliullin}},\ and\ \bibinfo {author} {\bibfnamefont
  {B.~J.}\ \bibnamefont {Kim}},\ }\href {https://doi.org/10.1038/nphys3322}
  {\bibfield  {journal} {\bibinfo  {journal} {Nature Physics}\ }\textbf
  {\bibinfo {volume} {11}},\ \bibinfo {pages} {462} (\bibinfo {year}
  {2015})}\BibitemShut {NoStop}%
\bibitem [{\citenamefont {Hogan}\ \emph {et~al.}(2015)\citenamefont {Hogan},
  \citenamefont {Yamani}, \citenamefont {Walkup}, \citenamefont {Chen},
  \citenamefont {Dally}, \citenamefont {Ward}, \citenamefont {Dean},
  \citenamefont {Hill}, \citenamefont {Islam}, \citenamefont {Madhavan},\ and\
  \citenamefont {Wilson}}]{Hogan2015PRL}%
  \BibitemOpen
  \bibfield  {author} {\bibinfo {author} {\bibfnamefont {T.}~\bibnamefont
  {Hogan}}, \bibinfo {author} {\bibfnamefont {Z.}~\bibnamefont {Yamani}},
  \bibinfo {author} {\bibfnamefont {D.}~\bibnamefont {Walkup}}, \bibinfo
  {author} {\bibfnamefont {X.}~\bibnamefont {Chen}}, \bibinfo {author}
  {\bibfnamefont {R.}~\bibnamefont {Dally}}, \bibinfo {author} {\bibfnamefont
  {T.~Z.}\ \bibnamefont {Ward}}, \bibinfo {author} {\bibfnamefont {M.~P.~M.}\
  \bibnamefont {Dean}}, \bibinfo {author} {\bibfnamefont {J.}~\bibnamefont
  {Hill}}, \bibinfo {author} {\bibfnamefont {Z.}~\bibnamefont {Islam}},
  \bibinfo {author} {\bibfnamefont {V.}~\bibnamefont {Madhavan}},\ and\
  \bibinfo {author} {\bibfnamefont {S.~D.}\ \bibnamefont {Wilson}},\ }\href
  {https://doi.org/10.1103/PhysRevLett.114.257203} {\bibfield  {journal}
  {\bibinfo  {journal} {Phys. Rev. Lett.}\ }\textbf {\bibinfo {volume} {114}},\
  \bibinfo {pages} {257203} (\bibinfo {year} {2015})}\BibitemShut {NoStop}%
\bibitem [{\citenamefont {Chen}\ \emph {et~al.}(2018)\citenamefont {Chen},
  \citenamefont {Schmehr}, \citenamefont {Islam}, \citenamefont {Porter},
  \citenamefont {Zoghlin}, \citenamefont {Finkelstein}, \citenamefont {Ruff},\
  and\ \citenamefont {Wilson}}]{Chen2018NatComm}%
  \BibitemOpen
  \bibfield  {author} {\bibinfo {author} {\bibfnamefont {X.}~\bibnamefont
  {Chen}}, \bibinfo {author} {\bibfnamefont {J.~L.}\ \bibnamefont {Schmehr}},
  \bibinfo {author} {\bibfnamefont {Z.}~\bibnamefont {Islam}}, \bibinfo
  {author} {\bibfnamefont {Z.}~\bibnamefont {Porter}}, \bibinfo {author}
  {\bibfnamefont {E.}~\bibnamefont {Zoghlin}}, \bibinfo {author} {\bibfnamefont
  {K.}~\bibnamefont {Finkelstein}}, \bibinfo {author} {\bibfnamefont
  {J.~P.~C.}\ \bibnamefont {Ruff}},\ and\ \bibinfo {author} {\bibfnamefont
  {S.~D.}\ \bibnamefont {Wilson}},\ }\href
  {https://doi.org/10.1038/s41467-017-02647-1} {\bibfield  {journal} {\bibinfo
  {journal} {Nat. Comm.}\ }\textbf {\bibinfo {volume} {9}},\ \bibinfo {pages}
  {103} (\bibinfo {year} {2018})}\BibitemShut {NoStop}%
\bibitem [{\citenamefont {Harris}\ \emph {et~al.}(1995)\citenamefont {Harris},
  \citenamefont {Feng}, \citenamefont {Birgeneau}, \citenamefont {Hirota},
  \citenamefont {Shirane}, \citenamefont {Hase},\ and\ \citenamefont
  {Uchinokura}}]{Harris1995_PRB}%
  \BibitemOpen
  \bibfield  {author} {\bibinfo {author} {\bibfnamefont {Q.~J.}\ \bibnamefont
  {Harris}}, \bibinfo {author} {\bibfnamefont {Q.}~\bibnamefont {Feng}},
  \bibinfo {author} {\bibfnamefont {R.~J.}\ \bibnamefont {Birgeneau}}, \bibinfo
  {author} {\bibfnamefont {K.}~\bibnamefont {Hirota}}, \bibinfo {author}
  {\bibfnamefont {G.}~\bibnamefont {Shirane}}, \bibinfo {author} {\bibfnamefont
  {M.}~\bibnamefont {Hase}},\ and\ \bibinfo {author} {\bibfnamefont
  {K.}~\bibnamefont {Uchinokura}},\ }\href
  {https://doi.org/10.1103/PhysRevB.52.15420} {\bibfield  {journal} {\bibinfo
  {journal} {Phys. Rev. B}\ }\textbf {\bibinfo {volume} {52}},\ \bibinfo
  {pages} {15420} (\bibinfo {year} {1995})}\BibitemShut {NoStop}%
\bibitem [{\citenamefont {Wilkens}\ and\ \citenamefont
  {M\"uller-Buschbaum}(1991)}]{Wilkens1991}%
  \BibitemOpen
  \bibfield  {author} {\bibinfo {author} {\bibfnamefont {J.}~\bibnamefont
  {Wilkens}}\ and\ \bibinfo {author} {\bibfnamefont {H.}~\bibnamefont
  {M\"uller-Buschbaum}},\ }\href@noop {} {\bibfield  {journal} {\bibinfo
  {journal} {Zeitschrift f\"ur anorganische und allgemeine Chemie}\ }\textbf
  {\bibinfo {volume} {592}},\ \bibinfo {pages} {79} (\bibinfo {year}
  {1991})}\BibitemShut {NoStop}%
\bibitem [{sup()}]{supp}%
  \BibitemOpen
  \href@noop {} {}\bibinfo {note} {See Supplemental Material at ? for
  additional data of the nuclear superlattice peaks from the x-ray scattering
  experiments.}\BibitemShut {Stop}%
\bibitem [{\citenamefont {Pelissetto}\ and\ \citenamefont
  {Vicari}(2002)}]{Pelissetto2002}%
  \BibitemOpen
  \bibfield  {author} {\bibinfo {author} {\bibfnamefont {A.}~\bibnamefont
  {Pelissetto}}\ and\ \bibinfo {author} {\bibfnamefont {E.}~\bibnamefont
  {Vicari}},\ }\href@noop {} {\bibfield  {journal} {\bibinfo  {journal}
  {Physics Reports}\ }\textbf {\bibinfo {volume} {368}},\ \bibinfo {pages} {549
  } (\bibinfo {year} {2002})}\BibitemShut {NoStop}%
\bibitem [{\citenamefont {Birgeneau}\ \emph {et~al.}(1977)\citenamefont
  {Birgeneau}, \citenamefont {Als-Nielsen},\ and\ \citenamefont
  {Shirane}}]{Birgeneau1977_PRB}%
  \BibitemOpen
  \bibfield  {author} {\bibinfo {author} {\bibfnamefont {R.~J.}\ \bibnamefont
  {Birgeneau}}, \bibinfo {author} {\bibfnamefont {J.}~\bibnamefont
  {Als-Nielsen}},\ and\ \bibinfo {author} {\bibfnamefont {G.}~\bibnamefont
  {Shirane}},\ }\href {https://doi.org/10.1103/PhysRevB.16.280} {\bibfield
  {journal} {\bibinfo  {journal} {Phys. Rev. B}\ }\textbf {\bibinfo {volume}
  {16}},\ \bibinfo {pages} {280} (\bibinfo {year} {1977})}\BibitemShut
  {NoStop}%
\bibitem [{\citenamefont {Birgeneau}\ \emph {et~al.}(1999)\citenamefont
  {Birgeneau}, \citenamefont {Kiryukhin},\ and\ \citenamefont
  {Wang}}]{Birgeneau_1999PRB}%
  \BibitemOpen
  \bibfield  {author} {\bibinfo {author} {\bibfnamefont {R.~J.}\ \bibnamefont
  {Birgeneau}}, \bibinfo {author} {\bibfnamefont {V.}~\bibnamefont
  {Kiryukhin}},\ and\ \bibinfo {author} {\bibfnamefont {Y.~J.}\ \bibnamefont
  {Wang}},\ }\href {https://doi.org/10.1103/PhysRevB.60.14816} {\bibfield
  {journal} {\bibinfo  {journal} {Phys. Rev. B}\ }\textbf {\bibinfo {volume}
  {60}},\ \bibinfo {pages} {14816} (\bibinfo {year} {1999})}\BibitemShut
  {NoStop}%
\bibitem [{\citenamefont {Shashidhar}\ \emph {et~al.}(1988)\citenamefont
  {Shashidhar}, \citenamefont {Ratna}, \citenamefont {Nair}, \citenamefont
  {Prasad}, \citenamefont {Bahr},\ and\ \citenamefont
  {Heppke}}]{Shashidhar1988_PRL}%
  \BibitemOpen
  \bibfield  {author} {\bibinfo {author} {\bibfnamefont {R.}~\bibnamefont
  {Shashidhar}}, \bibinfo {author} {\bibfnamefont {B.~R.}\ \bibnamefont
  {Ratna}}, \bibinfo {author} {\bibfnamefont {G.~G.}\ \bibnamefont {Nair}},
  \bibinfo {author} {\bibfnamefont {S.~K.}\ \bibnamefont {Prasad}}, \bibinfo
  {author} {\bibfnamefont {C.}~\bibnamefont {Bahr}},\ and\ \bibinfo {author}
  {\bibfnamefont {G.}~\bibnamefont {Heppke}},\ }\href
  {https://doi.org/10.1103/PhysRevLett.61.547} {\bibfield  {journal} {\bibinfo
  {journal} {Phys. Rev. Lett.}\ }\textbf {\bibinfo {volume} {61}},\ \bibinfo
  {pages} {547} (\bibinfo {year} {1988})}\BibitemShut {NoStop}%
\bibitem [{\citenamefont {Kim}\ \emph {et~al.}(2002)\citenamefont {Kim},
  \citenamefont {Revaz}, \citenamefont {Zink}, \citenamefont {Hellman},
  \citenamefont {Rhyne},\ and\ \citenamefont {Mitchell}}]{Kim2002_PRL}%
  \BibitemOpen
  \bibfield  {author} {\bibinfo {author} {\bibfnamefont {D.}~\bibnamefont
  {Kim}}, \bibinfo {author} {\bibfnamefont {B.}~\bibnamefont {Revaz}}, \bibinfo
  {author} {\bibfnamefont {B.~L.}\ \bibnamefont {Zink}}, \bibinfo {author}
  {\bibfnamefont {F.}~\bibnamefont {Hellman}}, \bibinfo {author} {\bibfnamefont
  {J.~J.}\ \bibnamefont {Rhyne}},\ and\ \bibinfo {author} {\bibfnamefont
  {J.~F.}\ \bibnamefont {Mitchell}},\ }\href
  {https://doi.org/10.1103/PhysRevLett.89.227202} {\bibfield  {journal}
  {\bibinfo  {journal} {Phys. Rev. Lett.}\ }\textbf {\bibinfo {volume} {89}},\
  \bibinfo {pages} {227202} (\bibinfo {year} {2002})}\BibitemShut {NoStop}%
\bibitem [{\citenamefont {Boseggia}\ \emph
  {et~al.}(2013{\natexlab{b}})\citenamefont {Boseggia}, \citenamefont
  {Springell}, \citenamefont {Walker}, \citenamefont {R\o{}nnow}, \citenamefont
  {R\"uegg}, \citenamefont {Okabe}, \citenamefont {Isobe}, \citenamefont
  {Perry}, \citenamefont {Collins},\ and\ \citenamefont
  {McMorrow}}]{Boseggia2013_PRL}%
  \BibitemOpen
  \bibfield  {author} {\bibinfo {author} {\bibfnamefont {S.}~\bibnamefont
  {Boseggia}}, \bibinfo {author} {\bibfnamefont {R.}~\bibnamefont {Springell}},
  \bibinfo {author} {\bibfnamefont {H.~C.}\ \bibnamefont {Walker}}, \bibinfo
  {author} {\bibfnamefont {H.~M.}\ \bibnamefont {R\o{}nnow}}, \bibinfo {author}
  {\bibfnamefont {C.}~\bibnamefont {R\"uegg}}, \bibinfo {author} {\bibfnamefont
  {H.}~\bibnamefont {Okabe}}, \bibinfo {author} {\bibfnamefont
  {M.}~\bibnamefont {Isobe}}, \bibinfo {author} {\bibfnamefont {R.~S.}\
  \bibnamefont {Perry}}, \bibinfo {author} {\bibfnamefont {S.~P.}\ \bibnamefont
  {Collins}},\ and\ \bibinfo {author} {\bibfnamefont {D.~F.}\ \bibnamefont
  {McMorrow}},\ }\href {https://doi.org/10.1103/PhysRevLett.110.117207}
  {\bibfield  {journal} {\bibinfo  {journal} {Phys. Rev. Lett.}\ }\textbf
  {\bibinfo {volume} {110}},\ \bibinfo {pages} {117207} (\bibinfo {year}
  {2013}{\natexlab{b}})}\BibitemShut {NoStop}%
\bibitem [{\citenamefont {Dhital}\ \emph {et~al.}(2013)\citenamefont {Dhital},
  \citenamefont {Hogan}, \citenamefont {Yamani}, \citenamefont {de~la Cruz},
  \citenamefont {Chen}, \citenamefont {Khadka}, \citenamefont {Ren},\ and\
  \citenamefont {Wilson}}]{Dhital2013}%
  \BibitemOpen
  \bibfield  {author} {\bibinfo {author} {\bibfnamefont {C.}~\bibnamefont
  {Dhital}}, \bibinfo {author} {\bibfnamefont {T.}~\bibnamefont {Hogan}},
  \bibinfo {author} {\bibfnamefont {Z.}~\bibnamefont {Yamani}}, \bibinfo
  {author} {\bibfnamefont {C.}~\bibnamefont {de~la Cruz}}, \bibinfo {author}
  {\bibfnamefont {X.}~\bibnamefont {Chen}}, \bibinfo {author} {\bibfnamefont
  {S.}~\bibnamefont {Khadka}}, \bibinfo {author} {\bibfnamefont
  {Z.}~\bibnamefont {Ren}},\ and\ \bibinfo {author} {\bibfnamefont {S.~D.}\
  \bibnamefont {Wilson}},\ }\href {https://doi.org/10.1103/PhysRevB.87.144405}
  {\bibfield  {journal} {\bibinfo  {journal} {Phys. Rev. B}\ }\textbf {\bibinfo
  {volume} {87}},\ \bibinfo {pages} {144405} (\bibinfo {year}
  {2013})}\BibitemShut {NoStop}%
\bibitem [{\citenamefont {Kubota}\ \emph {et~al.}(2015)\citenamefont {Kubota},
  \citenamefont {Tanaka}, \citenamefont {Ono}, \citenamefont {Narumi},\ and\
  \citenamefont {Kindo}}]{Kubota2015PRB}%
  \BibitemOpen
  \bibfield  {author} {\bibinfo {author} {\bibfnamefont {Y.}~\bibnamefont
  {Kubota}}, \bibinfo {author} {\bibfnamefont {H.}~\bibnamefont {Tanaka}},
  \bibinfo {author} {\bibfnamefont {T.}~\bibnamefont {Ono}}, \bibinfo {author}
  {\bibfnamefont {Y.}~\bibnamefont {Narumi}},\ and\ \bibinfo {author}
  {\bibfnamefont {K.}~\bibnamefont {Kindo}},\ }\href
  {https://doi.org/10.1103/PhysRevB.91.094422} {\bibfield  {journal} {\bibinfo
  {journal} {Phys. Rev. B}\ }\textbf {\bibinfo {volume} {91}},\ \bibinfo
  {pages} {094422} (\bibinfo {year} {2015})}\BibitemShut {NoStop}%
\bibitem [{\citenamefont {Chaloupka}\ and\ \citenamefont
  {Khaliullin}(2016)}]{Chaloupka2016PRB}%
  \BibitemOpen
  \bibfield  {author} {\bibinfo {author} {\bibfnamefont {J.~c.~v.}\
  \bibnamefont {Chaloupka}}\ and\ \bibinfo {author} {\bibfnamefont
  {G.}~\bibnamefont {Khaliullin}},\ }\href
  {https://doi.org/10.1103/PhysRevB.94.064435} {\bibfield  {journal} {\bibinfo
  {journal} {Phys. Rev. B}\ }\textbf {\bibinfo {volume} {94}},\ \bibinfo
  {pages} {064435} (\bibinfo {year} {2016})}\BibitemShut {NoStop}%
\bibitem [{\citenamefont {Ju}\ \emph {et~al.}(2013)\citenamefont {Ju},
  \citenamefont {Liu},\ and\ \citenamefont {Yang}}]{Ju2013PRB}%
  \BibitemOpen
  \bibfield  {author} {\bibinfo {author} {\bibfnamefont {W.}~\bibnamefont
  {Ju}}, \bibinfo {author} {\bibfnamefont {G.-Q.}\ \bibnamefont {Liu}},\ and\
  \bibinfo {author} {\bibfnamefont {Z.}~\bibnamefont {Yang}},\ }\href
  {https://doi.org/10.1103/PhysRevB.87.075112} {\bibfield  {journal} {\bibinfo
  {journal} {Phys. Rev. B}\ }\textbf {\bibinfo {volume} {87}},\ \bibinfo
  {pages} {075112} (\bibinfo {year} {2013})}\BibitemShut {NoStop}%
\bibitem [{\citenamefont {Wills}(2000)}]{Wills2000}%
  \BibitemOpen
  \bibfield  {author} {\bibinfo {author} {\bibfnamefont {A.}~\bibnamefont
  {Wills}},\ }\href@noop {} {\bibfield  {journal} {\bibinfo  {journal} {Physica
  B: Condensed Matter}\ }\textbf {\bibinfo {volume} {276-278}},\ \bibinfo
  {pages} {680 } (\bibinfo {year} {2000})}\BibitemShut {NoStop}%
\bibitem [{\citenamefont {Klein}\ \emph {et~al.}(2011)\citenamefont {Klein},
  \citenamefont {Rousse}, \citenamefont {Damay}, \citenamefont {Porcher},
  \citenamefont {Andr\'e},\ and\ \citenamefont {Terasaki}}]{Klein2011PRB}%
  \BibitemOpen
  \bibfield  {author} {\bibinfo {author} {\bibfnamefont {Y.}~\bibnamefont
  {Klein}}, \bibinfo {author} {\bibfnamefont {G.}~\bibnamefont {Rousse}},
  \bibinfo {author} {\bibfnamefont {F.}~\bibnamefont {Damay}}, \bibinfo
  {author} {\bibfnamefont {F.}~\bibnamefont {Porcher}}, \bibinfo {author}
  {\bibfnamefont {G.}~\bibnamefont {Andr\'e}},\ and\ \bibinfo {author}
  {\bibfnamefont {I.}~\bibnamefont {Terasaki}},\ }\href
  {https://doi.org/10.1103/PhysRevB.84.054439} {\bibfield  {journal} {\bibinfo
  {journal} {Phys. Rev. B}\ }\textbf {\bibinfo {volume} {84}},\ \bibinfo
  {pages} {054439} (\bibinfo {year} {2011})}\BibitemShut {NoStop}%
\bibitem [{\citenamefont {Liu}\ and\ \citenamefont
  {Khaliullin}(2019)}]{Liu2019}%
  \BibitemOpen
  \bibfield  {author} {\bibinfo {author} {\bibfnamefont {H.}~\bibnamefont
  {Liu}}\ and\ \bibinfo {author} {\bibfnamefont {G.}~\bibnamefont
  {Khaliullin}},\ }\href@noop {} {\bibfield  {journal} {\bibinfo  {journal}
  {Phys. Rev. Lett.}\ }\textbf {\bibinfo {volume} {122}},\ \bibinfo {pages}
  {057203} (\bibinfo {year} {2019})}\BibitemShut {NoStop}%
\bibitem [{\citenamefont {Porras}\ \emph {et~al.}(2019)\citenamefont {Porras},
  \citenamefont {Bertinshaw}, \citenamefont {Liu}, \citenamefont {Khaliullin},
  \citenamefont {Sung}, \citenamefont {Kim}, \citenamefont {Francoual},
  \citenamefont {Steffens}, \citenamefont {Deng}, \citenamefont {Sala},
  \citenamefont {Efimenko}, \citenamefont {Said}, \citenamefont {Casa},
  \citenamefont {Huang}, \citenamefont {Gog}, \citenamefont {Kim},
  \citenamefont {Keimer},\ and\ \citenamefont {Kim}}]{Porras2019PRB}%
  \BibitemOpen
  \bibfield  {author} {\bibinfo {author} {\bibfnamefont {J.}~\bibnamefont
  {Porras}}, \bibinfo {author} {\bibfnamefont {J.}~\bibnamefont {Bertinshaw}},
  \bibinfo {author} {\bibfnamefont {H.}~\bibnamefont {Liu}}, \bibinfo {author}
  {\bibfnamefont {G.}~\bibnamefont {Khaliullin}}, \bibinfo {author}
  {\bibfnamefont {N.~H.}\ \bibnamefont {Sung}}, \bibinfo {author}
  {\bibfnamefont {J.-W.}\ \bibnamefont {Kim}}, \bibinfo {author} {\bibfnamefont
  {S.}~\bibnamefont {Francoual}}, \bibinfo {author} {\bibfnamefont
  {P.}~\bibnamefont {Steffens}}, \bibinfo {author} {\bibfnamefont
  {G.}~\bibnamefont {Deng}}, \bibinfo {author} {\bibfnamefont {M.~M.}\
  \bibnamefont {Sala}}, \bibinfo {author} {\bibfnamefont {A.}~\bibnamefont
  {Efimenko}}, \bibinfo {author} {\bibfnamefont {A.}~\bibnamefont {Said}},
  \bibinfo {author} {\bibfnamefont {D.}~\bibnamefont {Casa}}, \bibinfo {author}
  {\bibfnamefont {X.}~\bibnamefont {Huang}}, \bibinfo {author} {\bibfnamefont
  {T.}~\bibnamefont {Gog}}, \bibinfo {author} {\bibfnamefont {J.}~\bibnamefont
  {Kim}}, \bibinfo {author} {\bibfnamefont {B.}~\bibnamefont {Keimer}},\ and\
  \bibinfo {author} {\bibfnamefont {B.~J.}\ \bibnamefont {Kim}},\ }\href
  {https://doi.org/10.1103/PhysRevB.99.085125} {\bibfield  {journal} {\bibinfo
  {journal} {Phys. Rev. B}\ }\textbf {\bibinfo {volume} {99}},\ \bibinfo
  {pages} {085125} (\bibinfo {year} {2019})}\BibitemShut {NoStop}%
\bibitem [{\citenamefont {Cao}\ \emph {et~al.}(1998)\citenamefont {Cao},
  \citenamefont {Bolivar}, \citenamefont {McCall}, \citenamefont {Crow},\ and\
  \citenamefont {Guertin}}]{Cao1998PRB}%
  \BibitemOpen
  \bibfield  {author} {\bibinfo {author} {\bibfnamefont {G.}~\bibnamefont
  {Cao}}, \bibinfo {author} {\bibfnamefont {J.}~\bibnamefont {Bolivar}},
  \bibinfo {author} {\bibfnamefont {S.}~\bibnamefont {McCall}}, \bibinfo
  {author} {\bibfnamefont {J.~E.}\ \bibnamefont {Crow}},\ and\ \bibinfo
  {author} {\bibfnamefont {R.~P.}\ \bibnamefont {Guertin}},\ }\href
  {https://doi.org/10.1103/PhysRevB.57.R11039} {\bibfield  {journal} {\bibinfo
  {journal} {Phys. Rev. B}\ }\textbf {\bibinfo {volume} {57}},\ \bibinfo
  {pages} {R11039} (\bibinfo {year} {1998})}\BibitemShut {NoStop}%
\bibitem [{\citenamefont {O'Neal}\ \emph {et~al.}(2019)\citenamefont {O'Neal},
  \citenamefont {Paul}, \citenamefont {al~Wahish}, \citenamefont {Hughey},
  \citenamefont {Blockmon}, \citenamefont {Luo}, \citenamefont {Cheong},
  \citenamefont {Zapf}, \citenamefont {Topping}, \citenamefont {Singleton},
  \citenamefont {Ozerov}, \citenamefont {Birol},\ and\ \citenamefont
  {Musfeldt}}]{ONeal2019_npjQM}%
  \BibitemOpen
  \bibfield  {author} {\bibinfo {author} {\bibfnamefont {K.~R.}\ \bibnamefont
  {O'Neal}}, \bibinfo {author} {\bibfnamefont {A.}~\bibnamefont {Paul}},
  \bibinfo {author} {\bibfnamefont {A.}~\bibnamefont {al~Wahish}}, \bibinfo
  {author} {\bibfnamefont {K.~D.}\ \bibnamefont {Hughey}}, \bibinfo {author}
  {\bibfnamefont {A.~L.}\ \bibnamefont {Blockmon}}, \bibinfo {author}
  {\bibfnamefont {X.}~\bibnamefont {Luo}}, \bibinfo {author} {\bibfnamefont
  {S.-W.}\ \bibnamefont {Cheong}}, \bibinfo {author} {\bibfnamefont {V.~S.}\
  \bibnamefont {Zapf}}, \bibinfo {author} {\bibfnamefont {C.~V.}\ \bibnamefont
  {Topping}}, \bibinfo {author} {\bibfnamefont {J.}~\bibnamefont {Singleton}},
  \bibinfo {author} {\bibfnamefont {M.}~\bibnamefont {Ozerov}}, \bibinfo
  {author} {\bibfnamefont {T.}~\bibnamefont {Birol}},\ and\ \bibinfo {author}
  {\bibfnamefont {J.~L.}\ \bibnamefont {Musfeldt}},\ }\href@noop {} {\bibfield
  {journal} {\bibinfo  {journal} {npj Quantum Materials}\ }\textbf {\bibinfo
  {volume} {4}},\ \bibinfo {pages} {48} (\bibinfo {year} {2019})}\BibitemShut
  {NoStop}%
\bibitem [{\citenamefont {Revelli}\ \emph {et~al.}(2019)\citenamefont
  {Revelli}, \citenamefont {Loo}, \citenamefont {Kiese}, \citenamefont
  {Becker}, \citenamefont {Fr\"ohlich}, \citenamefont {Lorenz}, \citenamefont
  {Moretti~Sala}, \citenamefont {Monaco}, \citenamefont {Buessen},
  \citenamefont {Attig}, \citenamefont {Hermanns}, \citenamefont {Streltsov},
  \citenamefont {Khomskii}, \citenamefont {van~den Brink}, \citenamefont
  {Braden}, \citenamefont {van Loosdrecht}, \citenamefont {Trebst},
  \citenamefont {Paramekanti},\ and\ \citenamefont
  {Gr\"uninger}}]{Revelli2019PRB}%
  \BibitemOpen
  \bibfield  {author} {\bibinfo {author} {\bibfnamefont {A.}~\bibnamefont
  {Revelli}}, \bibinfo {author} {\bibfnamefont {C.~C.}\ \bibnamefont {Loo}},
  \bibinfo {author} {\bibfnamefont {D.}~\bibnamefont {Kiese}}, \bibinfo
  {author} {\bibfnamefont {P.}~\bibnamefont {Becker}}, \bibinfo {author}
  {\bibfnamefont {T.}~\bibnamefont {Fr\"ohlich}}, \bibinfo {author}
  {\bibfnamefont {T.}~\bibnamefont {Lorenz}}, \bibinfo {author} {\bibfnamefont
  {M.}~\bibnamefont {Moretti~Sala}}, \bibinfo {author} {\bibfnamefont
  {G.}~\bibnamefont {Monaco}}, \bibinfo {author} {\bibfnamefont {F.~L.}\
  \bibnamefont {Buessen}}, \bibinfo {author} {\bibfnamefont {J.}~\bibnamefont
  {Attig}}, \bibinfo {author} {\bibfnamefont {M.}~\bibnamefont {Hermanns}},
  \bibinfo {author} {\bibfnamefont {S.~V.}\ \bibnamefont {Streltsov}}, \bibinfo
  {author} {\bibfnamefont {D.~I.}\ \bibnamefont {Khomskii}}, \bibinfo {author}
  {\bibfnamefont {J.}~\bibnamefont {van~den Brink}}, \bibinfo {author}
  {\bibfnamefont {M.}~\bibnamefont {Braden}}, \bibinfo {author} {\bibfnamefont
  {P.~H.~M.}\ \bibnamefont {van Loosdrecht}}, \bibinfo {author} {\bibfnamefont
  {S.}~\bibnamefont {Trebst}}, \bibinfo {author} {\bibfnamefont
  {A.}~\bibnamefont {Paramekanti}},\ and\ \bibinfo {author} {\bibfnamefont
  {M.}~\bibnamefont {Gr\"uninger}},\ }\href
  {https://doi.org/10.1103/PhysRevB.100.085139} {\bibfield  {journal} {\bibinfo
   {journal} {Phys. Rev. B}\ }\textbf {\bibinfo {volume} {100}},\ \bibinfo
  {pages} {085139} (\bibinfo {year} {2019})}\BibitemShut {NoStop}%
\end{thebibliography}%

\end{document}